\documentclass[11pt]{article}
\usepackage{CJK}
\usepackage{graphicx,psfrag}
\usepackage{fancyhdr}
\usepackage{amsmath,amsthm,amssymb}
\usepackage{mathrsfs}
\usepackage{multirow,booktabs}
\usepackage{indentfirst}
\usepackage{pifont}
\usepackage{color}
\usepackage{amsfonts}
\usepackage{url}
\usepackage{epstopdf}
\def\hDash{\bot\!\!\!\bot}

\newtheorem{theorem}{Theorem}[section]

\numberwithin{equation}{section}

\begin{document}

\title{ A projection-based adaptive-to-model test for regressions
\footnote{Lixing Zhu is a Chair professor of Department of Mathematics
at Hong Kong Baptist University, Hong Kong, China. He was supported by a grant from the
University Grants Council of Hong Kong, Hong Kong, China.}}
\author{Falong Tan$^{1,2}$, Xuehu Zhu$^3$ and Lixing Zhu$^2$
\\ $^1$ School of Mathematics and Statistics, Shen Zhen University, Shen Zhen, China\\
 $^2$ Department of Mathematics, Hong Kong Baptist University, Hong Kong\\
 $^3$ School of Mathematics, Xi'an Jiaotong University, Xi'an, China
}
\date{}
\maketitle

\renewcommand\baselinestretch{1.5}
{\small}

\begin{abstract}  A longstanding problem of existing
empirical process-based  tests for regressions is that when the number of covariates is greater than one, they either have no tractable limiting null distributions or are not omnibus.  To attack this problem, we in this paper propose a projection-based adaptive-to-model approach. When the hypothetical model is parametric single-index, the method can fully utilize the dimension reduction model structure under  the null hypothesis as if the covariate were one-dimensional such that the martingale transformation-based test  can  be asymptotically distribution-free.  Further,   the test can    automatically adapt to the underlying  model structure such that the test can be omnibus and thus  detect alternative models distinct from the hypothetical model at the fastest possible rate in hypothesis testing. The method is examined through simulation studied and  is illustrated by a real data analysis.

\end{abstract}
\newpage
\baselineskip=21pt

\newpage

\setcounter{equation}{0}
\section{Introduction}

Even when the dimension of covariates is moderate, dimensionality still causes data structure  not to be  visualized and thus makes regression modelling difficult.  Therefore, in regression analysis,  dimension reduction model structure is often used to  approximate underlying models. A typical example is the parametric single-index regression model:
\begin{equation}\label{1.1}
Y = g(\beta_0^{\top}X, \theta_0) + \varepsilon,
\end{equation}
where $Y$ is the response variable with the covariates $X \in \mathbb{R}^p $ , $g(\cdot)$ is a  known smooth function, $\beta_0 \in \mathbb{R}^p$ and $\theta_0 \in \mathbb{R}^d$ are the unknown regression parameter vectors,  $\varepsilon$ is the error term
with $E(\varepsilon|X)=0$ and the notation $\top$ denotes transposition.

It is necessary to  check the mis-specification of the regression function such that further regression analysis can be proceeded. Thus,  the saturated alternative model is considered:
 \begin{equation}\label{1.2}
Y = G(X) + \varepsilon,
\end{equation}
where $G(\cdot)$ denotes an unknown smooth function.
 There are several methods available to test the null hypothesis of model~(\ref{1.1}), which can be used for  more general hypothetical  parametric models. As this paper focuses on  dimension-reduction issue, we only briefly mention existing locally and globally smoothing tests and then give a more detailed comment on existing methods that are used to handle the curse of dimensionality.
Locally smoothing tests include H\"{a}rdle and Mammen (1993), Zheng (1996), Fan and Li ( 1996), Dette (1999), Fan and Huang (2001), Koul and Ni (2004) and Van Keilegom et al.(2008). In low-dimensional cases, this type of tests can be sensitive to high-frequent alternative models. However, these tests rely on nonparametric regression estimation and thus suffer severely from the curse of dimensionality. This is  because nonparametric regression estimation is very inefficient in high-dimensional scenarios. Guo et al. (2015) had detailed comments. Globally smoothing tests are nonparametric estimation free and particularly sensitive to low frequency alternative models and have better asymptotic behaviours. This is because they are the averages over empirical processes. Examples include  Stute (1997), Stute et. al. (1998a), Stute et. al. (1998b), Zhu (2003), Khmadladze and Koul (2004), Stute, Xu and Zhu (2008).  For more references, see the review paper by  Gonz\'{a}lez-Manteiga and Crujeiras (2013).  However, when the dimension is greater than $1$, they  are usually not asymptotically distribution-free and thus require Monte Carlo approximations such as the wild bootstrap to determine critical values. Stute et al. (1998a) is a typical reference for this type of tests.  

To attack this longstanding problem, there are several efforts in the literature to alleviate the curse of dimensionality. Guo, et al. (2015), as a first attempt in this field,  suggested a model adaptive test that can avoid the dimensionality problem largely, but still requires nonparametric estimation. Thus, the test has slower convergence rate than $1/\sqrt n$ and theoretically, cannot detect the alternatives only distinct from the null at this fastest possible rate in hypothesis testing. 
A commonly used and efficient idea is to construct tests that are based on projected covariates in lower dimensional space. Most of existing methods are inspired by the projection pursuit technique that was first proposed by Friedman and Stuetzle (1981), since it is essential to find one or a few directions along which the departures from hypothetical models can be easily detected. Escanciano (2006) and Lavergne and Patilea (2008, 2012) proposed tests that are based on projected covariates. Two earlier and relevant references are Zhu and An (1992)  and Zhu and Li (1998).  Zhu (2003)  and  Stute, Xu and Zhu (2008) used residual processes to construct tests that can also be regarded as dimension reduction type. These tests usually need to resort to Monte Carlo approximations  to determine critical values (e.g. Escanciano 2006; and Lavergne and Patilea 2008) though some of them are even asymptotically distribution-free such as Lavergne and Patilea (2012). This is either because of  intractability of the null distribution or because of  computational instability and complexity  caused by the computation over all projected covariates at all directions. A relevant reference about the computation issue is Wong et al. (1995).  Xia (2009) also proposed a projection-based test that however has no way to control type I error.

More specifically,  existing projection-based tests that involve residual-marked empirical processes are  either the supremum or integral over all  projected covariates $\{ a^{\top}X: a\in R^p  \, \, {\rm with} \, \, \|a\|=1  \}$ to form  Kolmogonov-Smirnov type or Cr\"amer-von Mises type statistics.  It is worthwhile to note that test statistics naturally involve all projections under both the null and alternative hypothesis. Although it is  reasonable and the omnibus property can also be guaranteed, the limiting null distributions are often intractable.   In contrast, Stute and Zhu (2002) simply used one projection $\beta_0^{\top}X$ and thus the test behaves like the one with one-dimensional covariate.
For  model~(\ref{1.1})   letting $\epsilon=Y- g(\beta_0^{\top}X, \theta_0)$, we have that  under the null hypothesis,
\begin{eqnarray*}
E(\epsilon|X)=0 \Rightarrow E[Y- g(\beta_0^{\top}X, \theta_0)] I(\beta_0^{\top}X \le u)=0 \quad {\rm for\ all} \quad u \in \mathbb{R},
\end{eqnarray*}
 the residual marked empirical process defined by Stute and Zhu (2002) is
\begin{eqnarray}\label{1.3}
R_{n}(u)= n^{-1/2}\sum_{i=1}^n [Y_i- g(\beta^{\top}_n X_i,\theta_n)]I(\beta^{\top}_n X_i \leq u),
\end{eqnarray}
where $\{(X_1, Y_1),\cdots (X_n, Y_n)\}$ denote an $i.i.d.$ sample from the distribution of $(X,Y)$, $\beta_n$ and $\theta_n$ are, under the null hypothesis, root-$n$ consistent estimators of $\beta$ and $\theta$, respectively. The martingale transformation can lead to an asymptotically distribution-free test (Stute et al. 1998a). However, {\it the test  obviously fails to be omnibus} (see the comment in Escanciano 2006) because the construction only uses the model structure under the null hypothesis. Guo et al. (2015) gave an example to explicitly illustrate this phenomenon.

%

%

The purpose of this paper is to construct a globally smoothing test that inherits the asymptotically distribution-free and dimension reduction properties of Stute and Zhu's (2002) test under the null hypothesis and the omnibus property of general projection-based tests under the alternative hypothesis such as Escanciano (2006) and Lavergne and Patilea (2008). To simultaneously achieve these two goals, we suggest an adaptive-to-model martingale transformation approach that can make the test  automatically adapt to the underlying model structure under the respective null and alternative hypothesis. 

To accommodate more general alternatives, we consider the following model:
\begin{equation}\label{1.4}
Y = G(B^{\top}X) + \epsilon,
\end{equation}
where $G$ is an unknown smooth function, $B$ is a $p \times q$ matrix with $q$ orthogonal columns for an unknown $q$ with $1\leq  q \leq p$ and  $E(\epsilon|X)=0$.
When $q=1$ and $B=\kappa\beta_0$ for some constant $\kappa$,  model (\ref{1.4}) becomes a semiparametric single-index model with the same index parameter as that in model~(\ref{1.1}). When $q=p$, model (\ref{1.4}) reduces to model~(\ref{1.2}) since $G(\cdot)$ is unknown and $G(X) = G(BB^{\top}X)  \equiv:  \tilde{G}(B^{\top}X)$.

This paper is organised as follows. Basic test construction is described in Section~2. As sufficient dimension reduction technique is crucial to implement the adaptive-to-model strategy for test construction, we also give a short review in this section.
In Section~3, we first present the asymptotic properties of the residual marked empirical process under the null hypothesis.
The martingale transformation-based innovative process is then discussed. After that, we investigate the properties of the process and its innovative process under the alternative hypothesis. In Section~4, the test statistic is presented and simulation results for small to moderate sample size are reported and a real data analysis is used as an illustration of application. Appendix contains technical proofs of the theoretical results.

\section{Projection-based adaptive-to-model empirical process}
\subsection{Basic  construction}
The null hypothesis now can be restated as
\begin{eqnarray*}
H_0:\ E(Y|X)= g(\beta_0^{\top}X, \theta_0)  \quad \rm for\ some\  \quad
\beta_0 \in \mathbb{R}^p,\ \theta_0 \in \Theta \subset \mathbb{R}^d
\end{eqnarray*}
and the alternative hypothesis is that for any $\beta \in \mathbb{R}^p$ , $\theta \in \mathbb{R}^d$ and a $ p\times q\ $  matrix $B$
\begin{eqnarray*}
H_1:\ E(Y|X)= G(B^{\top}X) \neq g(\beta^{\top}X,\theta),
\end{eqnarray*}
where  $G$ is unknown. Without loss of generality, assume $\beta_0$ is a linear combination of the columns of $B$. Recall $\epsilon=Y-g(\beta_0^{\top}X,\theta_0)$.
Under the null hypothesis, $q=1$ and $B=\kappa\beta_0$ for some constant $\kappa$, then we have $E(\epsilon|\beta_0^{\top}X)=E(\epsilon|B^{\top}X)=0$.
Under the alternative hypothesis $E(\epsilon|B^{\top}X)=G(B^{\top}X)-g(\beta_0^{\top}X, \theta_0) \neq 0$.
Thus, under the null hypothesis,
\begin{equation}\label{2.1}
E[(Y-g({\beta_0}^{\top}X, \theta_0))I(\beta_0^{\top}X \le u)]=0.
\end{equation}
According to Lemma~1 of Escanciaco (2006), Lemma~2.1 of Lavergne  and Patilea (2008) or a similar result in Zhu and Li (1998) that can be traced back to Zhu and An (1992), we have that,  under the alternative hypothesis, for an $\alpha \in \mathbb{R}^q$ whose first element is positive and $\| \alpha\|=1$
\begin{equation}\label{2.2}
E[(Y-g({\beta_0}^{\top}X, \theta_0))I(\alpha^{\top}B^{\top}X \le u)] \neq 0.
\end{equation}
Note that under the null and alternative hypothesis, we use respective $\beta_0^{\top}X$ and $\alpha^{\top}B^{\top}X$. It is clear that we cannot define two estimates separately according to null  and alternative hypothesis as we do not know the underlying model, while need an estimate $\hat{B}_n$ of $B$ that can adapt the underlying model: under the null $\hat{B}_n$ converges to a vector proportional to $\beta_0$ and under the alternative, to $B$. If this can be achieved, we can use the empirical version of the left hand side of (\ref{2.2}) to be the basis of a test statistic.
Let $\{(X_1,Y_1),\dots,(X_n,Y_n)\}$ be a sample with the same distribution as $(X,Y)$.
Thereby we propose an adaptive-to-model residual marked empirical process for checking model (\ref{1.1}) as follows:
\begin{equation}\label{2.3}
V_{n}(u,\hat{\alpha})= n^{-1/2}\sum_{i=1}^n[Y_i- g(\beta^{\top}_{n}X_i, \theta_n)]I(\hat{\alpha}^{\top} \hat{B}_n^{\top}X_i \leq u),
\end{equation}
\begin{equation}\label{2.4}
V_{n}(u)=\sup_{\| \hat{\alpha} \|=1,a_1\geq 0} V_{n}(u,\hat{\alpha})
\end{equation}
where $\hat{\alpha}^{\top}=(a_1,\dots,a_{\hat{q}}) \in \mathbb{R}^{\hat{q}}$,  $\hat{B}_n$ is a sufficient dimension reduction estimator of $B$ with an estimated structural dimension $\hat{q}$ of $q$, $\beta_n$ and $\theta_n$ are  respectively ordinary least squares estimators of $\beta_0$ and $\theta$.

It is clear that in order to have the model adaptation property of the process such that under the null hypothesis, $\sup_{\| \hat{\alpha} \|=1,a_1\geq 0} V_{n}(u,\hat{\alpha})$ is equal to $n^{-1/2}\sum_{i=1}^n[Y_i- g(\beta^{\top}_{n}X_i, \theta_n)]\\I( {\beta}_n^{\top}X_i \leq u)$ Stute and Zhu (2002) defined, we must have that under the null hypothesis, $\hat{q}$ and $\hat{B}$ converge to $1$ and $\kappa\beta_0$, respectively. Thus, we discuss their estimations.
%
\subsection{A review on discretization-expectation estimation}
To identify and estimate the number $q$ and the matrix $B$, we use a method of sufficient dimension reduction (SDR). There are several proposals available in the literature. Examples include sliced inverse regression (SIR,Li 1991), sliced average variance estimation (SAVE, Cook and Weisberg 1991), minimum average variance estimation (MAVE, Xia et.al. 2002), directional regression (DR, Li and Wang, 2007), likelihood acquired directions (LAD, Cook and Forzani, 2009) and average partial mean estimation (APME, Zhu et al. 2010b). In this section we briefly review discretization-expectation estimation (DEE, Zhu, et al. 2010b).  Since $G$ is unknown, for any $q \times q$ orthogonal matrix $C$, $G(B^{\top}X)$ can be rewritten as $\tilde{G}(C^{\top}B^{\top}X)$. This means $B$ is not identifiable in  model (\ref{1.4}). Thus,  SDR methodologies show that  we can only identify  $q$ base vectors in the central mean subspace $\mathcal{S}_{E(Y|X)}$ spanned by $B$ (see Cook (1998)), or precisely $BC$ for a $q\times q$ orthonormal matrix $C$. This can be achieved through identifying  $\mathcal{S}_{E(Y|X)}$. In the SDR theory, $\mathcal{S}_{E(Y|X)}$ is defined to be the intersection of all subspaces $\rm{span}(A)$ such that $Y \hDash E(Y|X)|A^{\top}X$ where  $\hDash$ means  statistical independence and $\rm{span}(A)$ means the subspace spanned by the columns of $A$.  The dimension of $\mathcal{S}_{E(Y|X)}$ is called the structural dimension, denoted as $d_{E(Y|X)}$. Therefore, under the null hypothesis(\ref{1.1}), $\mathcal{S}_{E(Y|X)}=\rm{span}(\beta_0/\|\beta_0\|)$ and $d_{E(Y|X)}=1$; while under the alternative (\ref{1.4}), $\mathcal{S}_{E(Y|X)}=\rm{span}(B)$ and $d_{E(Y|X)}=q$.
Similarly, the central subspace (Cook (1998)) denoted by $\mathcal{S}_{Y|X}$ is defined to be the intersection of all subspaces $\rm{span}(A)$ such that $Y \hDash X|A^{\top}X$. It is easy to see that $\mathcal{S}_{E(Y|X)} \subset \mathcal{S}_{Y|X}$. For simplicity, we assume $ \mathcal{S}_{E(Y|X)}=\mathcal{S}_{Y|X}$. A special case is that $\eta \hDash X$ in model (\ref{1.4}).

The basic procedure  of DEE is given below.
\begin{enumerate}
\item Define the discrete response variable $Z(t)= I\{Y \leq t\}$ where the indicator function $I\{Y \leq t\}$ take  value 1 if $Y \leq t$ and 0 otherwise.
\item Let $\mathcal{S}_{Z(t)|X}$ denote the central subspace of $Z(t)|X$ and $M(t)$ be a $ p\times  p$ positive semi-definite matrix such that $\rm{Span}\{M(t)\} =\emph{S}_{Z(t)|X}$.
\item Let $\tilde Y$ be an independent copy of $Y$ and $M=E\{M(\tilde Y)\}$. Theorem 1 in Zhu et al. (2010) asserts that $\rm{Span}(M) =\mathcal{S}_{Y|X}$ and $ B$ consists of the eigenvectors corresponding to the nonzero eigenvalues of $M$.
\item The estimator of the target matrix $M$ is given by
\begin{equation*}
M_{n}=\frac{1}{n}\sum^{n}_{i=1}M_n({y}_i),
\end{equation*}
where $M_n({y}_i)$ is the estimator of the matrix $M({y}_i)$ obtained by a chosen sufficient dimension reduction method such as SIR (Li 1991). Then an estimator $ B_n(q)$ of $B$ consists of the eigenvectors associated with the largest $q$ eigenvalues of $M_{n}$ when $q$ is given.
\end{enumerate}
According to Theorems~2 and  3 in Zhu et al. (2010), $ B_n(q)$ could achieve root-$n$ consistence to $B$. For more details the readers can refer to  Zhu et al. (2010).

\subsection{ Structural dimension  estimation}
As argued in Section~2.2, a consistency estimator of the structure dimension $q$ is required. Zhu et al.(2010a) suggested the BIC-type criterion to determine $q$, which is a modification of the method in Zhu et al. (2006).  Here, we suggest a minimum ridge-type eigenvalue ratio estimate (MRER) to determine the structure dimension $q$. Let  $\hat{\lambda}_{p} \leq \cdots \leq \hat{\lambda}_{1}$ denote the eigenvalues of the estimated matrix $M_{n}$ of $M$. Note that $q=dim\mathcal{S}_{Y|X}=rank(M)$.  Then the true structure dimension $q$ can be estimated by
\begin{equation}\label{2.5}
\hat{q}=\arg\min_{1\leq i \leq p}\left\{i: \frac{\hat{\lambda}^2_{i+1}+c}{\hat{\lambda}^2_i+c}\right\}.
\end{equation}
The algorithm of MRER is easy to implement. The following theorem shows that the consistency of MRER is adaptive to the underlying model.

\noindent{\bf Lemma 1 }\label{Lemma1}
{\it
Under the regularity conditions assumed by Zhu et al. (2010a), the estimator $\hat{q}$ of (\ref{2.5}) with $c=\log{n}/n$ satisfies that, as $ n\rightarrow \infty $,
\begin{itemize}
\item [(i)] under $H_0$, $Pr(\hat{q}=1)\rightarrow 1 $,
\item [(ii)] under $H_1$, $Pr(\hat{q}=q)\rightarrow 1 $.
\end{itemize}}
A justification of this lemma  can be found  in the Appendix.

\section{Main results}
\subsection{Basic properties of the process}
First, we consider the following process
\begin{equation*}
V^{0}_{n}(u,\alpha)= n^{-1/2}\sum_{i=1}^n[Y_i- g(\beta_0^{\top}X_i,\theta_0)]I(\alpha^{\top}B^{\top}X_i \leq u),
\end{equation*}
where  $\alpha=(a_1,a_2,\dots,a_q)^{\top}, a_1 \geq 0$ and $\|\alpha\|=1$. Denote $\sigma^2(v,\alpha)=Var(Y|\alpha^{\top}B^{\top}X=v)$ and $\psi(u,\alpha) =E[Var(Y|\alpha^{\top}B^{\top}X)I(\alpha^{\top}B^{\top}X \leq u)]$.
Under the null hypothesis,  $q=1,\alpha=1$ and $B=\kappa \beta_0$. Thus, we rewrite it as
\begin{eqnarray*}
\sigma^2(v) & \equiv:& \sigma^2(v,1)=Var(Y|\kappa\beta_0^{\top}X=v),\\
\psi(u)&\equiv:& \psi(u,1)=E[Var(Y|\kappa\beta_0^{\top}X)I(\kappa \beta_0^{\top}X \leq u)],\\
V^{0}_n(u)  &\equiv:&  V^{0}_{n}(u,1)= n^{-1/2}\sum_{i=1}^n[Y_i- g(\beta_0^{\top}X_i,\theta_0)]I(\kappa \beta_0^{\top}X_i \leq u).
\end{eqnarray*}
Obviously, $\psi(u)=\int^u_{-\infty}{\sigma^2(v)}dF_{\kappa\beta_0}(dv)$ where $F_{\kappa \beta_0}$ denotes the distribution of $\kappa\beta_0^{\top}X$.

Under the null hypothesis,
 Theorem~1.1 in Stute (1997) implies that
\begin{eqnarray}\label{3.1}
V^0_{n}(u) \longrightarrow V_{\infty}(u) \quad \rm in\ distribution
\end{eqnarray}
in the Skorohod space $D[-\infty, \infty)$ where $V_{\infty}$ is a continuous Gaussian process with mean zero and covariance kernel $K(u_1,u_2)=\psi (u_1 \wedge u_2).$

 To study the process $V_n(u,\hat{\alpha})$ proposed in (\ref{2.3}), we give some regularity conditions on the function $g(\beta^{\top}X,\theta)$ and the parameters in the following.
\begin{itemize}
\item[A1] Under $H_0$, we suppose that $(\beta_n, \theta_n)$ has a linear expansion
\begin{eqnarray*}
\sqrt{n}\left( \begin{array}{c}
\beta_n-\beta_0\\
\theta_n-\theta_0\\
\end{array} \right)
=\frac{1}{\sqrt{n}}\sum_{i=1}^nl(x_i,y_i,\beta_0,\theta_0)+o_p(1),
\end{eqnarray*}
where $l$ is a vector-valued function satisfying that
\begin{itemize}
\item [I] $E(l(X, Y,\beta_0, \theta_0))=0$;
\item [II] $L(\beta_0, \theta_0) = E(l(X, Y,\beta_0, \theta_0)l^{\top}(X, Y,\beta_0, \theta_0))$ is positive definite.
\end{itemize}
It is easy to see that the ordinary least squares estimator satisfies condition (A1).
\item [A2] The function $g(\beta^{\top}x, \theta)$ is continuously differentiable with respect to $(\beta, \theta)$ in some neighbourhood of $(\beta_0, \theta_0)$. The first-order partial derivatives
\begin{eqnarray*}
m(x, \beta, \theta) = \frac{\partial{g(\beta^{\top}x,\theta)}}{\partial{(\beta,\theta)}}=(m_1(x,\beta, \theta),\cdots,m_{p+d}(x,\beta, \theta))^{\top}
\end{eqnarray*}
satisfies that there exists a $\mu-$integrable function $K_0(x)$ such that
\begin{eqnarray*}
|m_j(x,\beta, \theta)|\leq K_0(x) \quad {\rm for\ all } \quad (\beta,\theta) \quad {\rm and } \quad 1 \leq j \leq p+d.
\end{eqnarray*}
where $\mu$ denotes the distribution of $X$.
\item [A3] Let $H(u,\beta)=E[Var(Y|X)I(\kappa\beta^{\top}X \leq u)]$ and we assume that $H(u,\beta)$ is uniformly continuous in $u$ at $\beta_0$.
\end{itemize}

\begin{theorem} \label{Theorem3.1}
 Under $H_0$ and Conditions A1-A2, we have in distribution
\begin{eqnarray*}
V_{n}(u) \longrightarrow V_{\infty}(u)-M(u)^{\top}V \equiv: V^1_{\infty}(u)
\end{eqnarray*}
where $V_{\infty}(u)$ has the same Gaussian process as given by (\ref{3.1}), the vector-valued function $M^{\top}=(M_1,M_2,\cdots,M_{p+d})$ is defined as
$$M_i(u)=E(m_i(X, \beta_0, \theta_0)I(\kappa \beta_0^{\top}X \leq u))$$
and $V$ is a $p+d-$dimensional normal vector with zero means and covariance matrix $L(\beta_0, \theta_0)$.
 \end{theorem}

Theorem \ref{Theorem3.1} is almost the same as Theorem 1 in Stute and Zhu(2002) except for the definition of $M$. In other words, under the null hypothesis, our process can have almost the same limiting property as that in Stute and Zhu(2002). Thus, a martingale transformation can be implemented. However, we also need to check whether it can be done under the alternative hypothesis. We then discuss the adaptive-to-model martingale transformation below.

\subsection{Adaptive-to-model martingale transformation}
 To have the model adaptation property of the process,  our idea is that under the null hypothesis, the process is only with $\beta_0^{\top}X$ (or proportional to it as $\kappa\beta_0^{\top}X$), and under the alternatives, the process is automatically with $\alpha^{\top}B^{\top}X$ such that the transformation is still implementable and the transformed process can capture the information of  alternative models.

First, motivate our method from  the transformation under the null hypothesis. Recall $M$ a vector-valued function on $\mathbb{R}$ and
$$ \psi (u) = \int^u_{-\infty}{\sigma^2(v)}dF_{\kappa \beta_0}(dv)$$
a nonnegative increasing function with $ \psi (-\infty)=0$.
Let $a=\frac{\partial{M}}{\partial{\psi}}$ stand for the Radon-Nikodym derivative of $G$ w.r.t. $\psi$, assuming that it exists.
Let
\begin{eqnarray*}
A(u)=\int_{u}^{-\infty}a(v)a^{\top}(v)\psi(dv)
=\int_{u}^{-\infty}a(v)a^{\top}(v)\sigma^2(v)dF_{\kappa \beta_0}(dv)
\end{eqnarray*}
be a $(d+p)\times (d+p)$ matrix.
Finally, we define the innovation process transformation as
\begin{eqnarray}\label{3.2}
(Tf)(z)=f(z)-\int^{z}_{-\infty}a^{\top}(u)A^{-1}(u)\left[\int_{u}^{\infty}a(v)f(dv)\right]\psi(du).
\end{eqnarray}
Here we suppose that $A(u)$ is non-singular and the process $f$ should be either bounded variation or Brownian motion.

Using the same arguments in the proofs of lemma 3.1 and 3.2 in Nikabadze and Stute(1997), we have the following two vital facts:
\begin{itemize}
\item [(i)] $T(G^{\top}V) \equiv 0,$
\item [(ii)] $TV_{\infty}=V_{\infty}$ in distribution.
\end{itemize}
Thus $TV_{\infty}$ is a centered Gaussian process with a covariance kernel $K(u_1,u_2)=\psi (u_1 \wedge u_2)$ which can be considered as the martingale part in the Doob-Meyer decomposition of $V_{\infty}^1$. See Stute et al. (1998b).

Note that $T$ relies on some unknown quantities and thus needs to be replaced by its empirical version.
To this end, let $g_1(t,\theta)=\frac{\partial g(t,\theta)}{\partial t}$ and $ g_2(t,\theta)=\frac{\partial g(t,\theta)}{\partial \theta}$,
then
\begin{eqnarray*}
  m(x,\beta_0,\theta_0)=(g_1(\beta_0^{\top}X,\theta_0)X^{\top}, g_2(\beta_0^{\top}X, \theta_0))^{\top}.
\end{eqnarray*}
Recall that $M(u)=E(m(X,\beta_0,\theta_0)I(\kappa \beta_0^{\top}X \leq u))$, then we obtain
\begin{eqnarray*}
 M(u)=\left( \begin{array}{c}
           E[g_1(\beta_0^{\top}X,\theta_0)X I(\kappa\beta_0^{\top}X \leq u)]  \\
           E[g_2(\beta_0^{\top}X,\theta_0)^{\top} I(\kappa\beta_0^{\top}X \leq u)]
             \end{array}
      \right)
     =\left( \begin{array}{c}
           \int_{-\infty}^{u}g_1(v/{\kappa},\theta_0)r(v) F_{\kappa \beta_0}(dv) \\
           \int_{-\infty}^{u}g_2(v/{\kappa},\theta_0)^{\top} F_{\kappa \beta_0}(dv)
             \end{array}
      \right),
\end{eqnarray*}
where $r(v)=E(X|\kappa \beta_0^{\top}X=v)$. It is easy to see that
\begin{eqnarray*}
a=\frac{\partial{M}}{\partial{\psi}}=\left( \begin{array}{c}
                                            g_1(v/{\kappa},\theta_0)r(v)/{\sigma^2(v)} \\
                                            g_2(v/{\kappa},\theta_0)^{\top}/{\sigma^2(v)}
                                            \end{array}
                                     \right),
\end{eqnarray*}
and
\begin{eqnarray*}
A(u)=\int_{u}^{-\infty}a(v)M^{\top}(dv)=\left( \begin{array}{c}
                                            \int_{u}^{-\infty}g_1(v/{\kappa},\theta_0)r(v)/{\sigma^2(v)}M^{\top}(dv) \\
                                            \int_{u}^{-\infty}g_2(v/{\kappa},\theta_0)^{\top}/{\sigma^2(v)}M^{\top}(dv)
                                            \end{array}
                                         \right).
\end{eqnarray*}

In a general nonparametric framework, there are no assumptions on $r$ and $\sigma$ except for smoothness, thus both the functions need to be estimated by some curve estimators. Here we adopt a standard Nadaraya-Watson estimator for $r$
\begin{eqnarray*}
r_{n}(v)=\frac{\sum_{i=1}^n X_i K(\frac{v-\hat{\alpha}^{\top}\hat{B}^{\top}_{n} X_i}{h})}{\sum_{i=1}^n K(\frac{v-\hat{\alpha}^{\top}\hat{B}^{\top}_{n} X_i}{h})}.
\end{eqnarray*}
For $\sigma^2$, note that  $\sigma^2(u)=E(\epsilon^2|\kappa\beta_0X=u)$ which can be replaced by
\begin{eqnarray*}
\sigma^2_{n}(u)=\frac{\sum_{i=1}^n \hat{\epsilon}^2_i K(\frac{u-\hat{\alpha}^{\top}\hat{B}^{\top}_{n} X_i}{h})}{\sum_{i=1}^n K(\frac{u-\hat{\alpha}^{\top}\hat{B}^{\top}_{n} X_i}{h})},
\end{eqnarray*}
where $K(\cdot)$ is a univariate kernel function and $h$ is a bandwidth. It is worth mentioning that we use $\hat{\alpha}^{\top}\hat{B}^{\top}_{n}X$ rather than $ \beta_n^{\top}X$ where $ \beta_n$ is the nonlinear least squares estimate of $\beta_0$ that was used by Stute and Zhu (2002). As $\hat q$ and $\hat B$ have the model adaptation property, in the following, we can derive the model adaptation property of the  transformed process.

Now we can respectively obtain the empirical versions $a_n, M_n$ and $ A_n$ of $a, M$ and $ A$:
\begin{eqnarray*}
  a_n(v) &=& \left( \begin{array}{c}
                     g_1(v/{\kappa_n},\theta_n)r_n(v)/{\sigma_n^2(v)} \\
                     g_2(v/{\kappa_n},\theta_n)^{\top}/{\sigma_n^2(v)}
                     \end{array}
             \right), \\
  M_n(u) &=& \left( \begin{array}{c}
                      \frac{1}{n} \sum_{i=1}^{n}g_1(\beta_n^{\top}X_i,\theta_n)X_i I(\hat{\alpha}^{\top} \hat{B}_n^{\top}X_i \leq u) \\
                      \frac{1}{n} \sum_{i=1}^{n}g_2(\beta_n^{\top}X_i,\theta_n)^{\top} I(\hat{\alpha}^{\top} \hat{B}_n^{\top}X_i \leq u)
                     \end{array}
             \right), \\
  A_n(u) &=& \frac{1}{n} \sum_{i=1}^{n} I(\hat{\alpha}^{\top} \hat{B}_n^{\top} X_i \geq u)
               \left( \begin{array}{c}
                       g_1(\hat{\alpha}^{\top} \hat{B}_n^{\top}X_i/{\kappa_n},\theta_n)r_n(\hat{\alpha}^{\top} \hat{B}_n^{\top}X_i)/{\sigma_n^2(\hat{\alpha}^{\top} \hat{B}_n^{\top}X_i)} \\
                       g_2(\hat{\alpha}^{\top} \hat{B}_n^{\top}X_i/{\kappa_n},\theta_n)^{\top}/{\sigma_n^2(\hat{\alpha}^{\top} \hat{B}_n^{\top}X_i)}
                      \end{array}
               \right)\\
               &&\times (g_1(\beta_n^{\top}X_i,\theta_n)X_i^{\top},g_2(\beta_n^{\top}X_i,\theta_n)).
\end{eqnarray*}
Replace $a, M$and $ A$ in (\ref{3.2}) by their empirical versions, we obtain the empirical version of $TV$:
\begin{eqnarray*}
(T_n V_n)(u,\hat{\alpha})&=& V_n(u,\hat{\alpha})- \int_{-\infty}^{u} a_n(v)^{\top} A_n^{-1}(v) \int_{v}^{\infty} a_n(z) V_n(dz)
                             \sigma_n^2(v)F_{\hat{\alpha}}(dv) \\
                       &=& \frac{1}{n^{1/2}} \sum_{i=1}^{n}[Y_i-g(\beta_n^{\top}X_i,\theta_n)]I(\hat{\alpha}^{\top}\hat{B}_n^{\top} X_i \leq u)-
                          \frac{1}{n^{3/2}}\sum_{i,j=1}^n I(\hat{\alpha}^{\top}\hat{B}_n^{\top} X_i \leq u)\\
                          && \times  [g_1(\hat{\alpha}^{\top}\hat{B}_n^{\top} X_i/{\kappa_n},\theta_n) r_n(\hat{\alpha}^{\top}\hat{B}_n^{\top}X_i)^{\top}, g_2(\hat{\alpha}^{\top} \hat{B}_n^{\top}X_i/{\kappa_n},\theta_n)] \\
                          && \times  A^{-1}_n(\hat{\alpha}^{\top}\hat{B}_n^{\top}X_i)
                           I(\hat{\alpha}^{\top}\hat{B}_n^{\top}X_j \geq \hat{\alpha}^{\top}\hat{B}_n^{\top}X_i)(Y_j-g(\beta^{\top}_{n}X_j,\theta_n))\\
                          && \times
                          \left( \begin{array}{c}
                            g_1(\hat{\alpha}^{\top} \hat{B}_n^{\top}X_i/{\kappa_n},\theta_n)r_n(\hat{\alpha}^{\top} \hat{B}_n^{\top}X_i)/{\sigma_n^2(\hat{\alpha}^{\top} \hat{B}_n^{\top}X_i)} \\
                            g_2(\hat{\alpha}^{\top} \hat{B}_n^{\top}X_i/{\kappa_n},\theta_n)^{\top}/{\sigma_n^2(\hat{\alpha}^{\top} \hat{B}_n^{\top}X_i)}
                                 \end{array}
                          \right)
\end{eqnarray*}
where $\kappa_n$ is the estimator of $\kappa$ and $F_{\hat{\alpha}}$ is the empirical distribution function of $\hat{\alpha}^{\top} \hat{B}_n^{\top} X_i,1\leq i \leq n.$

Theorems~\ref{Theorem3.2} and ~\ref{Theorem3.3} below show that   the resulting transformed process is adaptive to the underlying model.

\begin{theorem} \label{Theorem3.2}
 Let $\sigma^2_{n}(u)$ be a consistent estimator of $\sigma^2$ which is bounded away from zero. Under  the regularity conditions of Theorem (\ref{Theorem3.1}) and  $H_0$, we have in distribution
\begin{eqnarray*}
\sup_{\|\hat{\alpha}\|=1,a_1\geqslant0} T_nV_n(u,\hat{\alpha}) \longrightarrow V_{\infty}(u) \quad {\rm in\ disitribution\ in\ the\ space}\ D[-\infty, \infty).
\end{eqnarray*}
\end{theorem}

The result meets our expectation. Because of model adaptation, the supremum under the null  is actually over only one direction and the transformed process is a standard Gaussian process as proved by Stute and Zhu (2002).

When the distribution of $X$ is elliptically contoured, particularly spherically contoured such as normal distributions, the formulation of the transformation can be much simpler and thus the computation can be much easier.  Without loss of generality, consider spherically contoured distributions.   Suppose the regression function $g$ does not rely on the parameter $\theta$. Let $g'(\beta_0^{\top}x)$ be the derivative of $g(\cdot)$ about $\beta_0^{\top}x$. Thus we have $m(x,\beta_0)=g'(\beta_0^{\top}x)x$
and $$M(u)=E[g'(\beta_0^{\top}X)X I(\kappa \beta_0^{\top}X \leq u)]=\int^u_{-\infty}{g'(v/{\kappa})}r(v)dF_{\kappa \beta_0}(dv).$$
Therefore $a=\frac{\partial{M}}{\partial{\psi}}=\frac{g'(v/{\kappa})r(v)}{\sigma^2(v)}$ and $A(u)=\int_{u}^{-\infty}a(v)M^{\top}(dv)=\int_{u}^{-\infty} \frac{g'(v/{\kappa})r(v)}{\sigma^2(v)}M^{\top}(dv).$

Let $\Gamma$ be any orthogonal matrix with the first row $\beta_0^{\top}/\|\beta_0\|$, thus the first component of $\Gamma X$ is $\beta_0^{\top}X/\|\beta_0\|$. Since the conditional distribution of the other components of $\Gamma X$ given the first is still spherical, these conditional expectations are zero. Therefore,
\begin{eqnarray*}
M(u)&=&\Gamma^{\top}Eg'(\beta_0^{\top}X) \Gamma X I(\kappa \beta_0^{\top}X \leq u)\\
&=&\frac{\beta_0}{\|\beta_0\|^2} Eg'(\beta_0^{\top}X) \beta_0^{\top}X I(\kappa \beta_0^{\top}X \leq u) \\
&=& \frac{\beta_0}{\|\beta_0\|^2} \int_{-\infty}^{u} [g'(v/{\kappa})v/{\kappa}]F_{\kappa \beta_0}(dv).
\end{eqnarray*}
Thus we obtain
\begin{eqnarray*}
  a(v) &=& \frac{\beta_0}{\|\beta_0\|^2} \frac{g'(v/{\kappa})v/{\kappa}}{\sigma^2(v)},\\
  A(u) &=& \frac{\beta_0 \beta_0^{\top}}{\|\beta_0\|^4} \int_{u}^{\infty}\frac{[g'(v/{\kappa})v/{\kappa}]^2}{\sigma^2(v)}F_{\kappa \beta_0}(dv).
\end{eqnarray*}
The matrix $A(x)$ is singular  with rank $1$. To derive relevant asymptotic  results as those in  Theorem \ref{Theorem3.2},  denote
\begin{eqnarray*}
h(v) = Eg'(\beta_0^{\top}X) \beta_0^{\top}X I(\kappa \beta_0^{\top}X \leq v) \quad \rm{and} \quad  V_1 = \frac{\beta_0^{\top}V}{\|\beta_0\|^2}.
\end{eqnarray*}
Therefore $M^{\top}(u)V=h(u)V_1$ and the conclusion of Theorem \ref{Theorem3.1} would be rewritten as
$$V_{n} \longrightarrow V_{\infty}-hV_1  \quad \rm{in\ distribution}$$
The new $a$ and $A$ now become  real-valued:
$$a=\frac{\partial h}{\partial \psi} \quad {\rm and} \quad A(u)=\int_{u}^{\infty} a^2(v)\sigma^2(v)F_{\kappa \beta_0}(dv).$$
Similarly, we can also obtain the empirical analogues $A_n$ and $T_n$ of $A$ and $T$ respectively.
Therefore Theorem \ref{Theorem3.2} can be applied with these new functions in the results.

It is noteworthy to point out that convergence in $D[-\infty, \infty)$ means convergence in $D[-\infty, u]$ for any finite $u$. Since the transformation involves the inverses of $A(u)$, Our test statistic would yield instabilities in the distributional behaviors for large values of $u$. Thus all the processes should be constrained in  proper subsets of the real line. In practice, we would consider $T_nV_n$ on a given quantile of $\kappa_n\beta_n^{\top}X_i$, $1\leq i \leq n$. See Stute and Zhu (2002). Now we give the result to show that the transformed process can automatically adapt to the alternative models such that a constructed test can detect them.

\subsection{The properties under the alternative hypothesis}
 Consider the following sequence of  alternatives converging to the null hypothesis
\begin{eqnarray}\label{3.4}
H_{1n}:\ Y_n=g(\beta_0^{\top}X,\theta_0)+ C_n G(B^{\top}X)+\eta
\end{eqnarray}
where $E(\eta|X)=0$, and $\beta_0$ is the linear combination of the columns of $B$. When $C_n$ is a fixed constant, the model is under global alternative equivalent to model~(\ref{1.4}). If it tends to zero, the models are under local alternatives.
We  give the asymptotic property of the estimator $\hat{q}$ of the structure dimension $q$ under the local alternatives. The following lemma shows that when $C_n$ goes to zero quickly, $\hat q$ is not equal to $q$, while to $1$. In other words, $\hat q$ is an inconsistent estimate of $q$.

\noindent{\bf Lemma 2 }\label{Lemma2} {\it Under the local alternative $H_{1n}$ and the same conditions in Lemma 1 with $C_n = 1/\sqrt{n}$, the estimator $\hat{q}$ of (\ref{2.5}) with $c=\log{n}/n$ satisfies that $P(\hat{q}=1)\rightarrow 1$, as $ n\rightarrow \infty $. }

To derive the asymptotic properties of $\sup_{\|\hat{\alpha}\|=1,a_1\geqslant0} T_nV_n(u,\hat{\alpha})$, under the local alternative $H_{1n}$, we need an additional condition:
\begin{itemize}
\item [ A3] \begin{eqnarray*}
\sqrt{n}\left( \begin{array}{c}
                  \beta_n-\beta_0\\
                  \theta_n-\theta_0\\
                \end{array}
        \right)
=\gamma + \frac{1}{\sqrt{n}}\sum_{i=1}^nl(x_i, y_i,\beta_0,\theta_0)+o_p(1),
\end{eqnarray*}
\end{itemize}
where $\gamma$ is some constant vector and the vector-valued function $l$ is the same as given in (A1). Under $H_{1n}$ with $C_n=\frac{1}{\sqrt{n}}$, condition (A3) is satisfied for the nonlinear least squares estimate. See Lemma~3 in Guo et al.(2015).

Now we  derive the limits of $sup_{\|\hat{\alpha}\|=1,a_1 \geq 0}T_nV_{n}(x,\hat{\alpha})$ under the global alternative and local alternatives.
\begin{theorem} \label{Theorem3.3}
Under the regularity conditions A1-A3, we have\\
(i) under the global alternative $H_1$ that is equivalent to $H_{1n}$ with fixed $C_n$,
\begin{eqnarray*}
\frac{1}{\sqrt{n}} \sup_{\|\hat{\alpha}\|=1,a_1 \geq 0} T_nV_{n}(u,\hat{\alpha}) \to L(u);
\end{eqnarray*}
where $L(u)$ is some nonzero process. That is, $\sup_{\|\hat{\alpha}\|=1,a_1 \geq 0} T_nV_{n}(u,\hat{\alpha})$ diverges to infinity at the rate of $1/\sqrt n$. \\
(ii) under the local alternatives $H_{1n}$ with $C_n=1/{\sqrt{n}}$,
\begin{eqnarray*}
\sup_{\|\hat{\alpha}\|=1,a_1 \geq 0} T_nV_n(u,\hat{\alpha}) \rightarrow V_{\infty}(u)+\int_{-\infty}^{u} \{H_0(v)-a(v)^{\top} A^{-1}(v) [W_1(v)+W_2(v)]\sigma^2(v)\}F_{\kappa \beta_0}(dv)
\end{eqnarray*}
where $H_0(v)=E[G(B^{\top}X)|\kappa \beta_0^{\top}X=v], W_1(v)=\int_{v}^{\infty} a(z)H_0(z)F_{\kappa\beta_0}(dz)$ and
$W_2(v)$ have a normal distribution with mean zero and variance
$$\sigma_2^2(v)= \int_{v}^{\infty}\sigma^2(u)a(u)a(u)^{\top} F_{\kappa\beta_0}(du).$$

\end{theorem}
Based on Theorems~\ref{Theorem3.2} and \ref{Theorem3.3}, we can derive the asymptotic properties of  functionals of $\sup_{\|\hat{\alpha}\|=1,a_1 \geq 0} T_nV_{n}(u,\hat{\alpha})$ over all $u$. The resulting test statistic is defined in the next section.

\section{Numerical studies}
\subsection{Test statistics in practical use}  The test statistic is a functional of $T_{n}V_{n}$. In this paper, we consider the Cr\"amer$-$von Mises statistic of the form
\begin{eqnarray}\label{test}
CW^2_{n}=\int^{x_0}_{-\infty} \sup_{\|\hat{\alpha}\|=1,a_1\geq 0}(T_{n}V_{n}(u, \hat \alpha))^2 F_n(d\, u),
\end{eqnarray}
where $F_n$ is the empirical distribution function of $\frac{\beta_n}{\|\beta_n\|} X_i$, $1\leq i \leq n$. Note that we take $\kappa_n=1/{\|\beta_n\|}$ here. Using Theorem~\ref{Theorem3.2} and the continuous mapping theorem we obtain that, under the null hypothesis
\begin{eqnarray*}
CW_{n}^2 \longrightarrow  \int^{x_0}_{-\infty} B^2(\psi(u))F_{\kappa \beta_0}(du) \quad {\rm in\ disitribution}.
\end{eqnarray*}
where $B(u)$ is a standard Brownian motion.
To obtain a distribution-free limit for our test, note that
\begin{eqnarray*}
\frac{1}{\psi(x_0)^2} \int^{x_0}_{-\infty} B^2(\psi(u))\sigma^2(u) F_{\kappa \beta_0}(du)= \int_{0}^{1}B(u)^2 du \quad {\rm in\ disitribution},
\end{eqnarray*}
thus we consider
\begin{eqnarray*}
W^2_{n}=\frac{1}{\psi_n(x_0)^2}\int^{x_0}_{-\infty} \sup_{\|\hat{\alpha}\|=1,a_1\geqslant0}(T_{n}V_{n}(u, \hat \alpha))^2\sigma_n^2F_n(d\, u),
\end{eqnarray*}
where $\psi_n(u)=\frac{1}{n}\sum_{i=1}^{n}(Y_i-g(\beta_n^{\top}X_i))^2I(\frac{\beta_n}{\|\beta_n\|}X_i \leq u)$ is the estimator of $\psi(u)$ and $\sigma_n^2$ can be any consistent estimator of the conditional variance $\sigma^2$ defined in subsection 3.1. Therefore we have
\begin{eqnarray*}
W_{n}^2 \longrightarrow  \int^{1}_{0} B^2(u)du \quad {\rm in\ disitribution}.
\end{eqnarray*}

If the regression model is homoscedastic, then $\sigma^2$ is a constant and we can estimate it by
\begin{eqnarray*}
\sigma^2_n=\frac{1}{n}\sum_{i=1}^n [Y_i-g(\beta^{\top}_nX_i,\theta_n)]^2.
\end{eqnarray*}
Under the null hypothesis $\psi(x_0)=\sigma^2F_{\kappa\beta}(x_0)$ and it can be estimated by $\sigma^2_nF_{n}(x_0)$, thus ${W}^2_{n}$ becomes
\begin{eqnarray*}
{W}^2_{n} = \frac{1}{\sigma^2_nF_{n}^2(x_0)}\int^{x_0}_{-\infty} \sup_{\|\hat{\alpha}\|=1,a_1\geq 0}(T_{n}V_{n}(u, \hat \alpha))^2F_n(d\,u).
\end{eqnarray*}

For ease of comparison,we  give four examples in the following. For $x_0$, as Stute and Zhu (2002) did, we choose the $99\%$ quantile of $F_n$ in the numerical examples.


\subsection{Numerical examples}
In this subsection we conduct some simulations to show the performance of the distributional approximations for small to moderate sample size. We make a comparison with Guo et al.(2015)'s dimension reduction model-adaptive test $T_n^{GWZ}$ that is based on SIR-based DEE,  Stute et al.(1998a)'s test $T_n^{SGP}$ that determines critical values  by  the wild bootstrap.
We design four representative examples. The first is to confirm that the proposed test that can be regarded as an extension of Stue and Zhu's (2002) test is omnibus. The second includes both high-frequency and low-frequency model such as we can compare with an adaptive-to-model test that is based on locally smoothing approach.  The third includes models with higher  structural dimension and the fourth is also used to check the influence of dimensionality. The significant level is set to be $\alpha=0.05$, and the reported results are the average of $2000$ replications. In all models, the value of  $a=0$ corresponds to the null hypothesis and $a\neq 0$ to the alternatives.

$Example$ 1. The data are generated from the model
\begin{eqnarray*}
  Y =\frac{1}{4}\exp(2\beta_1^{\top}X)+a\beta_2^{\top}X+\varepsilon;
\end{eqnarray*}
Here we consider two cases: $p=3,\beta_1=(1,0,0)^{\top},\beta_2=(0,1,0)^{\top}$ and $p=4,\beta_1=(1,1,0,0)^{\top}/\sqrt{2},\beta_2=(0,0,1,1)^{\top}/\sqrt{2}$. In both cases, $n=50, 100$, $X$ follows the standard multivariate normal distribution $N(0,I_p)$ and $\varepsilon$ is from $N(0,1)$. Note that under the alternatives, we have $E(Y-\exp(\frac{5}{4}\beta_1^{\top}X)|\beta_1^{\top}X)=0$.
The results in Figure~1 obviously show that $T_n^{SZ}$ fails to work while $W_n^2$ performs very well.
 $$\rm{Figure \ 1 \ about \ here}$$

Now we consider the comparison between local and global smoothing tests and between the adaptive-to-model method and the classical method.\\

$Example$ 2. Consider
\begin{eqnarray*}
  H_{11}: Y&=& \beta_0^{\top}X+a \cos(\frac{\pi}{2} \beta_0^{\top}X)+\varepsilon; \\
  H_{12}: Y&=& \beta_0^{\top}X+\frac{1}{4}a \exp(\beta_0^{\top}X)+\varepsilon;\\
  H_{13}: Y&=& \beta_0^{\top}X+\frac{1}{2}a  (\beta_0^{\top}X)^2+\varepsilon.
\end{eqnarray*}
 where $p=8,\beta_0=(1,1,\dots,1)/{\sqrt{p}}$, $X=(X_1,X_2,\dots,X_p)^{\top}$ which is independent with $\varepsilon$. The central mean subspaces $\mathcal{S}_{E(Y|X)}$ has the structural dimension $1$ which means $B=\beta_0$ under both the null and alternative hypothesis. The predictors $x_i, i=1,\dots, n$ are  i.i.d. from two distributions: multivariate normal distributions $N(0,I_p)$ and $N(0,\Sigma)$ with $\Sigma=(1/2^{|i-j|})_{p\times p}$ so as to check the influence of correlation between the covariates. The errors $\varepsilon_i$'s are drawn independently from $N(0,1)$. The first is a high-frequency model and the others are low-frequent.

 The empirical sizes and powers of the three tests  are presented in Tables~1 and 2. We can see that both $T_n^{GWZ}$ and $W_n^2$ control the size very well even for the small size of $n=50$.  $T_n^{SGP}$ seems slightly more conservative with higher empirical size than $0.05$.
 For the first model,  $T_n^{GWZ}$ has relatively higher power than $W_n^2$ and $T_n^{SGP}$ have, especially in the correlated case with covariance matrix $\Sigma$.
For the models $H_{12}$ and $H_{13}$, both $W_n^2$ and $T_n^{SGP}$ are more powerful. These results further confirm that locally smoothing method performs better for high frequency models and globally smoothing method works better for low frequency models. The comparison also shows that $W_n^2$ is more robust against the underlying correlation  structure of the predictors than $T_n^{SGP}$ in both significance level maintainance and power performance.

$$ \rm{Table \ 1-2 \ about \ here}$$

To further investigate the performance of the proposed test, we consider the following model whose  structural dimension $q$ is greater than 1 under the alternatives. In this simulation, we  show the the advantage of our test in alleviating the dimensionality problem when compared to a locally smoothing test proposed by Zheng (1996) which can be regarded as one of the representative locally smoothing  methods.\\

 $Example$ 3. The data are generated from the model
 \begin{eqnarray*}
 H_{31}:Y&=&\beta_1^{\top}X+a(\beta_2^{\top}X)^2+\varepsilon;\\
 H_{32}:Y&=&\beta_1^{\top}X+a\exp\{-(\beta_2^{\top}X)^2\}+\varepsilon;
 \end{eqnarray*}
where $\beta_1=(\underbrace{1,\dots,1}_{p/2},0,\dots,0)^{\top}/\sqrt{p/2}$ and $\beta_2=(0,\dots,0,\underbrace{1,\dots,1}_{p/2})/\sqrt{p/2}$.
In this example, the predictors $x_i,i=1,\dots,n$ are i.i.d from  multivariate normal distribution $N(0,I_p)$ and $N(0,\Sigma)$ and $\varepsilon_i$'s are from the standard univariate normal distribution $N(0,1)$. In each case we have two cases with $p=2$ and $p=8$ respectively.


The simulation results are presented in Tables~3 and 4. When $p=2$, we can see that Zheng(1996)'s test $T_n^{ZH}$ can maintain the significance level occasionally, but usually, the empirical sizes are lower than $0.05$. In contrast,  $W_n^2$ works much better  even for $n=50$. For the empirical power, both $T_n^{ZH}$ and $W_n^2$ have high power. But the power of $W_n^2$ grows slightly faster as $a$ increases. When the dimension $p$ is $8$, the situation becomes very different. The empirical size of $T_n^{ZH}$ is far away from the significance level and its power becomes much lower than that in the $p=2$ case. Nevertheless, our test $W_n^2$ is much less unaffected by the dimensionality increasing than $T_n^{ZH}$. These phenomena validate the theoretical results that locally smoothing tests suffer from the dimensionality curse that causes slower convergence rate to their limits under the null and slower divergence rate to infinity under the alternative than globally smoothing tests.
 $$\rm{Table \ 3-4 \ about \ here}$$

In the following example, we consider a nonlinear null model against alternative models with higher structural dimensions. A more comprehensive comparison is made with Zheng(1996)'s test $T_n^{ZH}$,  Guo et al.(2015)'s test $T_n^{GWZ}$ and our test $W_n^2$.

$Example$ 4. Consider the following models
\begin{eqnarray*}
  H_{41}: Y&=&\exp(\frac{1}{2}X_1)+aX_2^3+\varepsilon;  \\
  H_{42}: Y&=&\exp(\frac{1}{2}X_1)+a\{X_2^3+\cos(\pi X_3)+ X_4 \}+\varepsilon;\\
  H_{43}: Y&=&\exp(\frac{1}{2}X_1)+a \{X_2^3+\cos(\pi X_3)+ X_4-|X_5|+X_6^2+ X_7 \times X_8 \}+ \varepsilon;
\end{eqnarray*}
where $(X_1,\dots,X_p)$ is independent of $\varepsilon$ and follows the standard multivariate normal distribution $N(0,I_p)$ with $p=4$ or $8$. Let $\beta_i$ be the unit vector with the $i$-th component $1$, $i=1, \dots, p$ and $a=0, 0.2, 0.4, \dots, 1$. When $a\neq 0$,  the structural dimension $q=2, B=(\beta_1,\beta_2)$ for $H_{41}$; $q=4, B=(\beta_1,\beta_2,\beta_3,\beta_4)$ for $H_{42}$ and $q=8, B=(\beta_1,\beta_2,\dots,\beta_8)$ for $H_{43}$. Note that under the alternatives, the models $H_{42}$ and $H_{43}$ do not have dimension reduction structure for $p=4$ and $p=8$ respectively. Thus, this can be used to further check the usefulness of the model adaptation method. The simulation results are reported in Figure~2.
 $$\rm{Figure \ 2 \ about \ here}$$

From this figure, we can see that when $p=4$, the performance of our test is slightly better than the other two competitors. However, when $p=8$, Zheng's test $T_n^{ZH}$ behaves much worse than $W_n^2$ and $T_n^{GWZ}$. This again indicates that  the dimensionality is a severe issue for the locally smoothing test without model adaptation as the adaptive-to-model test $T_n^{GWZ}$ can also work well though it is also locally smoothing-based.  For model $H_{42}$ with $p=4$ and model $H_{43}$ with $p=8$, $W_n^2$ and $T_n^{GWZ}$ still work well in the power performance, even though the model has no dimension reduction structure when $a\not =0$. Further,  globally smoothing-based test procedure shows its advantage as our test $W_n^2$ can outperform $T_n^{GWZ}$ when both are of the model adaptation property.

\subsection{Real data analysis}
This data set is studied 
to understand the various self-noise mechanisms. The data set is available at {UCI} Machine Learning Repository \url{https://archive.ics.uci.edu/ml/datasets/Airfoil+Self-Noise}. There are 1503 observations on one output variable: Scaled sound pressure level $Y$ (in decibels) and five input variables: Frequency $X_1$ (in Hertzs), Angle of attack $X_2$ (in degrees), Chord length $X_3$ (in meters), Free-stream velocity $X_4$ (in meters per second) and Suction side displacement thickness $X_5$ (in meters). For easy interpretation, all variables are standardized separately. To establish a regression relationship between $Y$ and 5 covariates $X=(X_1, \cdots, X_5)$, we try simple model first. When the dimension reduction is applied, we find that $Y$ may be conditionally independent of $X$ given a projected covariate $\beta_1^{\top}X$ in which the direction $\beta_1$ is searched by DEE. The scatter plot in Figure~3 shows a seemly linear relationship.
 $$\rm{Figure \ 3 \ about \ here}$$
To further explore the exhaustive search of projected covariables, we use the second porjected covariable searched by DEE, and then scatter plot of $Y$ against $(\beta_1^{\top}X, \beta_2^{\top}X)$ is presented in Figure~4.
 $$\rm{Figure \ 4 \ about \ here}$$
We can see clearly that the second direction $\beta_2$ is not necessary to use as the curves along the second direction could be almost identical. In other words, the projection of the data onto the space $\beta_1^{\top}X$ contains almost all information of model structure. Thus, we use a linear model to fit the data where the direction $\hat{\beta}_1
^{\top}=( -0.6323, -0.4339, -0.5339, 0.2386, \\-0.2644)$. To test whether the linear model is adequate, we use our test. The value of the test statistic $W_n^2=7.5322$ and the $p$-value is about $0$. Hence we need to further explore a possible model. When we use a polynomial to fit the data, a cubic polynomial of $\hat{\beta}_1^{\top}X$ may be appropriate. The fitted curve is added into the scatter plot. See  Figure~5.
 $$\rm{Figure \ 5 \ about \ here}$$

 The model is as follows:
\begin{eqnarray*}
  Y &=& \theta_1+\theta_2(\beta^{\top}X)+\theta_3(\beta^{\top}X)^2+\theta_4(\beta^{\top}X)^3+ \varepsilon.
\end{eqnarray*}
The value of the test statistic $W_n^2=0.1596$ and the p-value is $0.70$. Thus this model is plausible.

\section{Discussions}
In this paper, we propose a projection-based test that is based on residual marked empirical process and an adaptive-to-model martingale transformation. Compared to existing projection-based tests, the new test have the asymptotically distribution-free property under the null hypothesis and the omnibus property under the alternative hypothesis. This method is now for hypothetical models with dimension reduction structure. It is of great interest to investigate the application or extension of the method to hypothetical models without such kind of structure. The research is ongoing.

\section{Appendix}

\textbf{Proof of Lemma~1}. Under the regularity conditions given by Zhu et. al. (2010), Theorem 2 therein asserts that $M_{n}-M=O_p(n^{-1/2})$. Therefore, following the analogous argument of Zhu and Ng (1995) or Zhu and Fang (1996), we have $\hat{\lambda}_{i}- \lambda_{i} =O_p(n^{-1/2})$ for $i=1,\cdots, p$.

(I) Under $H_0$,  since $dim(\mathcal{S}_{Y|X})=1$, we have $\lambda_1>0,\lambda_l=0$ for any $l>1$. Therefore, $\hat{\lambda}^2_{1}=\lambda^2_{1}+O_p(n^{-1/2})$ and $\hat{\lambda}^2_{l}=O_p(n^{-1}), l=2,\dots, p$.
Hence, for any $l>1$,
\begin{eqnarray*}
\frac{\hat{\lambda}^2_{2}+c_n}{\hat{\lambda}^2_{1}+c_n}
&=&\frac{c_n+O_p(1/n)}{\lambda^2_1+c_n + O_p(1/\sqrt{n})} \rightarrow 0,\\
\frac{\hat{\lambda}^2_{l+1}+c_n}{\hat{\lambda}^2_{l}+c_n} &=&
\frac{c_n+O_p(1/n)}{c_n+O_p(1/n)}\rightarrow 1.
\end{eqnarray*}
Therefore, the minimizer $\hat q=1$ with a probability going to $1$.

(II) Under the alternative $H_1$, $dim(\mathcal{S}_{Y|X})=q$, we have $\lambda_l>0$ and $\hat{\lambda}^2_{l} = \lambda^2_{l}+O_p(1/\sqrt{n})$  for $l=1,\dots,q$  and $\hat{\lambda}^2_{l} = O_p(1/n)$ for $l=q+1,\dots, p$.  Hence, for $l<q$
\begin{eqnarray*}
\frac{\hat{\lambda}^2_{q+1}+c_n}{\hat{\lambda}^2_{q}+c_n}-\frac{\hat{\lambda}^2_{l+1}+c_n}{\hat{\lambda}^2_{l}+c_n}
&=&\frac{c_n+O_p(1/n)}{\lambda^2_{q}+c_n + O_p(1/\sqrt{n})}-\frac{\lambda^2_{l+1}+c_n+ O_p(1/\sqrt{n})}{\lambda_{l}^2+c_n+ O_p(1/\sqrt{n})}\\
&\rightarrow & -\frac{\lambda^2_{l+1}}{\lambda^2_{l}} < 0.
\end{eqnarray*}
For $l > q$
\begin{eqnarray*}
\frac{\hat{\lambda}^2_{q+1}+c_n}{\hat{\lambda}^2_{q}+c_n}-\frac{\hat{\lambda}^2_{l+1}+c_n}{\hat{\lambda}^2_{l}+c_n}
&=&\frac{c_n+O_p(1/n)}{\lambda^2_q+c_n + O_p(1/\sqrt{n})}-\frac{c_n+O_p(1/n)}{c_n+O_p(1/n)} \rightarrow  -1 < 0.
\end{eqnarray*}
Therefore, we can conclude that $Pr(\hat{q}= q)\longrightarrow 1$. \hfill$\Box$

\textbf{Proof of Theorem ~\ref{Theorem3.1}. } Under the null hypothesis, $Pr(\hat{q}= 1)\rightarrow 1$. Thus we can only work on the event $\hat{q}=1$ as the probability of the event $\hat{q} \not=1$ tends to $0$. Therefore $\hat{\alpha}=1$ and $V_n(u)=V_n(u,\hat{\alpha})$. Decompose the term $V_{n}(u)$ as follows
\begin{eqnarray*}
V_n(u)=V_n(u,\hat{\alpha}) &=& n^{-1/2}\sum_{i=1}^n\{Y_i- g(\beta^{\top}_{n}X_i, \theta_{n})\}I(\hat{B}^{\top}_{n}X_i \leq u)\\
                           &=&  n^{-1/2} \sum_{i=1}^n\{Y_i- g(\beta^{\top}_{n}X_i, \theta_{n})\}I(\kappa \beta_0^{\top}X_i \leq u)+\\
                           &&n^{-1/2} \sum_{i=1}^n\{Y_i- g(\beta^{\top}_{n}X_i, \theta_{n})\}[I(\hat{B}^{\top}_{n}X_i \leq u)-I(\kappa \beta_0^{\top}X_i \leq u)]\\
                           &\equiv:& V_{1n}+V_{2n}
\end{eqnarray*}
where $\kappa=1/\|\beta_0\|$.
Following the analogous argument of Theorem 1 in Stute and Zhu (2002), we obtain $V_{1n} \longrightarrow V_{\infty}-M(u)^{\top}V \equiv V^1_{\infty}$ and $V_{2n}$ tends to zero uniformly in $u$. \hfill$\Box$

\textbf{Proof of Lemma ~2}.
Using the same notations as in the proof of Lemma 1. Following the analogous argument for proving Theorem~2 in Guo et al.(2015), we obtain $M_{n}-M=O_p(n^{-1/2})$, therefore $\hat{\lambda}_{i}- \lambda_{i} =O_p(n^{-1/2})$ for $i=1,\cdots, p$. Note that $\lambda_l=0$ for any $l>1$. The proof is concluded from  the exact arguments for proving Lemma~1. \hfill$\Box$

\textbf{Proof of Theorem~\ref{Theorem3.2}}. Again, we only work on the event  $\hat q=1$ as $q=1$ under the null hypothesis. Denote $V_{n}^2(u)=n^{-1/2}\sum_{i=1}^{n}[Y_i-g(\beta_n^{\top}X_i, \theta_n)]I(\hat{B}_n^{\top}X_i \leq u)$. Under the null, $\hat{\alpha}=1$ and $\sup_{\|\hat{\alpha}\|=1,a_1 \geq 0}T_nV_n(u,\hat{\alpha})=T_nV_{n}^2(u)$. More explicitly,
$$T_nV_{n}^2(u)=V_{n}^2(u)-\int^{u}_{-\infty}a_n(v)^{\top} A^{-1}_n(v)[\int_{v}^{\infty}a_n(z) V_{n}^2(dz)] \sigma_n^2(v) F_{1n}(dv).$$
Let $V_{n}^1(u)=n^{-1/2}\sum_{i=1}^{n}[Y_i-g(\beta_0^{\top}X_i,\theta_0)]I(\hat{B}_n^{\top}X_i \leq u)$. Then
$$T_nV_{n}^1(u)=V_{n}^1(u)-\int^{u}_{-\infty}a_n(v)^{\top} A^{-1}_n(v)[\int_{v}^{\infty}a_n(z) V_{n}^1(dz)] \sigma_n^2(v) F_{1n}(dv).$$
Here $F_{1n}$ is the empirical distribution function of $\hat{B}_n^{\top}X_i, 1\leq i \geq n$.
Similarly as the arguments for proving Lemma 3.2 and Theorem 1.3 in Stute et al.(1998b), we obtain that
$$T_nV_{n}^2(u)=T_nV_{n}^1(u)+o_p(1),$$
$$T_nV_{n}^1(u)=TV_{n}^1(u)+o_p(1).$$
Recalling that under $H_0$ we have $V_n^0(u)=n^{-1/2}\sum_{i=1}^{n}[Y_i-g(\beta_0^{\top}X_i,\theta_0)]I(\kappa\beta_0^{\top}X_i \leq u)$, then
\begin{eqnarray*}
TV_n^0(u)-TV_{n}^1(u)&=&V_n^0(u)-V_{n}^1(u) \\
&&-\int_{-\infty}^{u} a(v)^{\top} A^{-1}(v) \int_{v}^{\infty} a(z) V_n^0(dz) \sigma^2(v) F_{\kappa\beta_0}(dv)\\
&&+\int_{-\infty}^{u} a(v)^{\top} A^{-1}(v) \int_{v}^{\infty} a(z) V_n^1(dz) \sigma^2(v) F_{\kappa\beta_0}(dv).
\end{eqnarray*}
 Using the same  proof  for Theorem~1 in Stute and Zhu(2002), we obtain that $TV_n^0-TV_{n}^1=o_p(1)$ uniformly in $u$. Therefore Lemma 3.3 in Stute et al.(1998b) gives our result.  \hfill$\Box$

\textbf{Proof of Theorem~\ref{Theorem3.3}}. (I) First we consider the global alternative hypothesis. Under the alternative $H_1$, Lemma~1 asserts that $Pr(\hat{q}= q)\to 1$, thus we work on the event $\hat{q}=q, \hat{\alpha}=\alpha=(a_1,\dots,a_q)^{\top}$.
Therefore on this event, $\sup_{\|\hat{\alpha}\|=1,a_1 \geq 0}T_nV_n(u,\hat{\alpha})=\sup_{\|\alpha\|=1,a_1 \geq 0}T_nV_n(u,\alpha)$. Denote $\delta_n=(\beta_n^{\top},\theta_n^{\top})^{\top}$ and $\delta_0=(\beta_0^{\top},\theta_0^{\top})^{\top}$.
According to White (1981), we have $\sqrt{n}(\delta_n-\tilde{\delta}_0)=O_p(1)$ where $\tilde{\delta}_0$ may not be equal to the true value $\delta_0$ under the null hypothesis.

Denote $$V_{n}^1(u,\alpha)=n^{-1/2}\sum_{i=1}^{n}[Y_i-g(\tilde{\beta}_0^{\top}X_i,\tilde{\theta}_0 )]I(\alpha^{\top}\hat{B}_n^{\top}X_i \leq u),$$
$$V_{n}^2(u,\alpha)=n^{-1/2}\sum_{i=1}^{n}[Y_i-g(\tilde{\beta}_0^{\top}X_i, \tilde{\theta}_0)]I(\alpha^{\top} B^{\top}X_i \leq u).$$
Similarly as the proof for Theorem 3.2, we obtain that
$$n^{-1/2}[T_nV_n(u,\alpha)-T_nV_{n}^1(u,\alpha)]=o_p(1),$$
$$n^{-1/2}[T_nV_n^1(u,\alpha)-TV_{n}^2(u,\alpha)]=o_p(1).$$
Here
\begin{eqnarray*}
n^{-1/2}TV_{n}^2(u,\alpha)&=&\frac{1}{n}\sum_{i=1}^{n}[Y_i-g(\tilde{\beta}_0^{\top}X_i,\tilde{\theta}_0)]I(\alpha^{\top} B^{\top}X_i \leq u)\\
&&-\frac{1}{\sqrt{n}} \int_{-\infty}^{u} a_1(v,\alpha)^{\top} A_1^{-1}(v,\alpha) \int_{v}^{\infty} a_1(z,\alpha) V_n^2(dz)\sigma_1^2(v,\alpha) F_{\alpha}(dv)
\end{eqnarray*}
Note that
$$\frac{1}{\sqrt{n}}\int_{v}^{\infty} a_1(z,\alpha) V_n^2(dz)=\frac{1}{n}\sum_{i=1}^{n}I(\alpha^{\top}B^{\top}X_i \geq v)a_1(\alpha^{\top}B^{\top}X_i,\alpha)[Y_i-g(\tilde{\beta}_0^{\top}X_i,\tilde{\theta}_0)]. $$
Then we derive that under $H_1$,
$$n^{-1/2}TV_{n}^2(u,\alpha) \to \int_{-\infty}^{u}H(v,\alpha)F_{\alpha}(dv) -\int_{-\infty}^{u} a_1(v,\alpha)^{\top} A_1^{-1}(v,\alpha) a_2(v,\alpha)\sigma_1^2(v,\alpha) F_{\alpha}(dv),$$
where $F_{\alpha}$ is the distribution function of $\alpha^{\top}B^{\top}X$ and
\begin{eqnarray*}
 H(v,\alpha)&=&E(G(B^{\top}X)-g(\tilde{\beta}_0^{\top}X,\tilde{\theta}_0)|\alpha^{\top}B^{\top}X=v),\\
 \sigma_1^2(v,\alpha)&=&E[(G(B^{\top}X)-g(\tilde{\beta}_0^{\top}X,\tilde{\theta}_0))^2+\varepsilon^2|\alpha^{\top}B^{\top}X=v],\\
 a_1(v,\alpha)&=& \{g_1(v/{\kappa_1},\tilde{\theta}_0)E(X^{\top}|\alpha^{\top}B^{\top}X=v)/{\sigma_1^2(v,\alpha)},g_2(v/{\kappa_1},\tilde{\theta}_0)
 /{\sigma_1^2(v,\alpha)}\}^{\top},\\
 A_1(v,\alpha)&=&E\{I(\alpha^{\top}B^{\top}X \geq v)a_1(\alpha^{\top}B^{\top}X,\alpha) (g_1(\tilde{\beta_0}^{\top}X,\tilde{\theta}_0)X^{\top},g_2(\tilde{\beta_0}^{\top}X,\tilde{\theta}_0))\},\\
 a_2(v,\alpha)&=&E\{I(\alpha^{\top}B^{\top}X \geq v) a_1(\alpha^{\top}B^{\top}X,\alpha) (G(B^{\top}X)-g(\tilde{\beta_0}^{\top}X,\tilde{\theta}_0))\}.
\end{eqnarray*}
Hence we conclude that
$$ n^{-1/2}T_nV_{n}(u,\alpha) \to \int_{-\infty}^{u}H(v,\alpha)F_{\alpha}(dv) -\int_{-\infty}^{u} a_1(v,\alpha)^{\top} A_1^{-1}(v,\alpha) a_2(v,\alpha)\sigma_1^2(v,\alpha) F_{\alpha}(dv). $$
Therefore
$$\frac{1}{\sqrt{n}}\sup_{\|\hat{\alpha}\|=1,a_1 \geq 0}T_nV_{n}(u,\hat{\alpha})
 \to \rm{some\ nonzero\ process}. $$
 The resulting test statistic converges to infinity at the rate of $O(1/n).$

(II) Under the local alternatives $H_{1n}$, Lemma 2 asserts that $P(\hat{q}=1) \to 1$ as $n \to \infty$,  thus we also consider the event $\hat{q}=1$.
 Denote $$V_{n}^2(u)=n^{-1/2}\sum_{i=1}^{n}[Y_i-g(\beta_n^{\top}X_i,\theta_n)]I(\hat{B}_n^{\top}X_i \leq u).$$
 Therefore $\hat{\alpha}=1$, $\hat B_n$ is a vector and
 $\sup_{\|\hat{\alpha}\|=1,a_1 \geq 0}T_nV_n(u,\hat{\alpha})=T_nV_{n}^2(u)$.
 Let $$V_{n}^1(u)=n^{-1/2}\sum_{i=1}^{n}[Y_i-g(\beta_0^{\top}X_i,\theta_n)]I(\hat{B}_n^{\top}X_i \leq u).$$
 Following the analogous argument for proving Theorem~3.3 we obtain that
 $$T_n V_{n}^2(u)=T_n V_{n}^1(u)+o_p(1),$$
 $$T_n V_{n}^1(u)=TV_{n}^1(u)+o_p(1),$$
 $$TV_n^0(u)=TV_{n}^1(u)+o_p(1).$$
 To finish the proof, it remains to derive the limit of $TV_n^0(u)$.
 Recall that  $V_n^0(u)=n^{-1/2}\sum_{i=1}^{n}[Y_i-g(\beta_0^{\top}X_i,\theta_0)]I(\kappa\beta_0^{\top}X_i \leq u)$ and
 $$TV_n^0(u)=V_n^0(u)-\int_{-\infty}^{u}a(v)^{\top}A^{-1}(v) \int_{v}^{\infty} a(z) V_n^0(dz) \sigma^2(v)F_{\kappa\beta_0}(dv).$$
 Under the local alternative $H_{1n}$,
 $$V_n^0(u)=\frac{1}{n}\sum_{i=1}^{n}G(B^{\top}X_i)I(\kappa\beta_0^{\top}X_i \leq u)+\frac{1}{\sqrt{n}}\sum_{i=1}^{n}\varepsilon_i I(\kappa\beta_0^{\top}X_i \leq u).$$
 Similarly as the proof for Theorem~1.1 in Stute(1997), we have
 $$V_n^0(u) \to V_{\infty}(u)+\int_{-\infty}^{u} H_0(v)F_{\kappa\beta_0}(dv),$$
 where $H_0(v)=E[G(B^{\top}X)|\kappa \beta_0^{\top}X=v].$

 For the second term in $TV_n^0(u)$, note that
 \begin{eqnarray*}
 \int_{v}^{\infty}a(z)V_n^0(dz)
 &=&\frac{1}{\sqrt{n}}\sum_{i=1}^{n}I(\kappa\beta_0^{\top}X_i \geq v) a(\kappa\beta_0^{\top}X_i) [Y_i-g(\beta_0^{\top}X_i,\theta_0)]\\
 &=&\frac{1}{n}\sum_{i=1}^{n}I(\kappa\beta_0^{\top}X_i \geq v) a(\kappa\beta_0^{\top}X_i) G(B^{\top}X_i)\\
 &&+\frac{1}{\sqrt{n}}\sum_{i=1}^{n}I(\kappa\beta_0^{\top}X_i \geq v) a(\kappa\beta_0^{\top}X_i) \varepsilon_i.
 \end{eqnarray*}
 Therefore,
 \begin{eqnarray*}
   \int_{v}^{\infty}a(z)V_n^0(dz) \rightarrow W_1(v)+ W_2(v)  \quad  \rm{in \ distribution}
 \end{eqnarray*}
 where $W_1(v)=\int_{v}^{\infty} a(z)H_0(z)F_{\kappa\beta_0}(dz)$ and $W_2(v)$ have a normal distribution with mean zero and variance
 $$\sigma_2^2(v)= \int_{v}^{\infty}\sigma^2(u)a(u)a(u)^{\top} F_{\kappa\beta_0}(du).$$
 Here $\sigma^2(u)=E(\varepsilon^2|\kappa\beta_0^{\top}X=u)$.
 Hence we can conclude that
 $$TV_n^0(u) \rightarrow V_{\infty}(u)+\int_{-\infty}^{u} \{H_0(v)-a(v)^{\top} A^{-1}(v) [W_1(v)+W_2(v)]\sigma^2(v)\}F_{\kappa \beta_0}(dv).$$  \hfill$\Box$

\newpage

\leftline{\large\bf References}

\begin{description}
\item Cook, R. D. (1998). {\it Regression Graphics: Ideas for Studying Regressions Through Graphics. } {New York: Wiley.}

\item Cook, R. D. and Weisberg, S. (1991). Discussion of ¡°Sliced inverse regression for dimension reduction,¡± by K. C. Li. {\it Journal of the American Statistical Association}, {\bf 86}, 316-342.

\item  Cook, R. D. and Forzani, L. (2009). Likelihood-based sufficient dimension reduction. {\it Journal of the American Statistical Association}, {\bf 104}, 197-208.

\item Dette, H. (1999). A consistent test for the functional form of a regression based on a difference of variance estimates. {\it The Annals of Statistics}. {\bf 27}, 1012-1050.

\item Escanciano, J. C.(2006). A consistent diagnostic test for regression models using projections. {\it Econometric Theory}, {\bf 22}, 1030-1051.

\item Eubank, R. L., Li, C. S. and Wang, S. (2005). Testing lack-of-fit of parametric regression models using nonparametric regression techniques. {\it Statistica Sinica}, {\bf 15}, 135-152.

\item Fan, J. Q. and Huang, L. S. (2001). Goodness-of-fit tests for parametric regression models, {\it Journal of the American Statistical Association}, {\bf 96}, 640-652.

\item Fan, Y. and  Li, Q. (1996). Consistent model specication tests: omitted variables and semiparametric functional forms. {\it Econometrica}, {\bf 64}, 865-890.

\item Fan, J., Zhang, C. and Zhang, J. (2001) Generalized likelihood ratio statistics and Wilks phenomenon. {\it The Annals of Statistics}, {\bf 29}, 153-193.

\item Friedman, J. H. and Stuetzle, W. (1981) Projection pursuit regression. {\it Journal of the American Statistical Association}, {\bf 76}, 817-823.

\item Guo, X., Wang, T. and Zhu, L. X. (2015). Model checking for generalized linear models: a dimension-reduction model-adaptive approach. {\it Journal of the Royal Statistical Society: Series B}, 

\item Gonz\'{a}lez-Manteiga, W. and Crujeiras, R. M. (2013). An updated review of Goodness-of-Fit tests for regression models. {\it TEST}, {\bf 22}, 361-411.

\item Khmadladze, E. V. and Koul, H. L. (2004) Martingale transforms goodness-of-fit tests in regression models. {\it The Annals of Statistics}, {\bf 37}, 995-1034

\item H\"{a}rdle, W. and  Mammen, E. (1993). Comparing nonparametric versus parametric regression fits. {\it The Annals of Statistics}, {\bf 21}, 1926-1947.

\item Hart, J. (1997) {\it Nonparametric smoothing and lack-of-fit tests.} {Springer, Berlin.}

\item Khmaladze, E V. (1982). Martingale Approach in the Theory of Goodness-of-fit Tests. {\it Theory of Probability $\&$ Its Applications},  {\bf 26}, 240-257.

\item Khmaladze, E V. and Koul, H. L. (2004). Martingale transforms goodness-of-fit tests in regression models. {\it The Annals of Statistics}, {\bf 32}, 995-1034.

\item Koul, H. L. and Ni, P. P. (2004). Minimum distance regression model checking. {\it Journal of Statistical Planning and Inference}, {\bf 119}, 109-141.

\item Lavergne, P. and Patilea, V. (2008). Breaking the curse of dimensionality in non parametric testing. {\it Journal of Econometrics}, {\bf 143}, 103-122.

\item Lavergne, P. and Patilea, V. (2012). One for all and all for one: regression checks with many regressors. {\it Journal of Business \& Economic Statistics}, {\bf 30}, 41-52.

\item Li, K. C. (1991). Sliced inverse regression for dimension reduction, {\it Journal of the American Statistical Association}, {\bf 86}, 316-327.

\item Li, B. and Wang, S. (2007). On directional regression for dimension reduction. {\it Journal of the American Statistical Association}, {\bf 102}, 997-1008.

\item Li, B., Wen, S. Q. and Zhu, L. X. (2008). On a Projective Resampling method for dimension reduction with multivariate responses. {\it Journal of the American Statistical Association.} {\bf 103}, 1177-1186.

\item Stute, W. (1997). Nonparametric model checks for regression. {\it The Annals of Statistics.} {\bf 25}, 613-641.

\item Stute, W., Gonz¡äales-Manteiga, W. and Presedo-Quindimil, M. (1998a). Bootstrap approximation in model checks for regression. {\it Journal of the American Statistical Association.}, {\bf 93}, 141-149.

\item Stute, W., Thies, S. and Zhu, L. X. (1998b). Model checks for regression: An innovation process approach. {\it The Annals of Statistics}. {\bf 26}, 1916-1934.

\item Stute,W., Xu, W. L. and Zhu, L. X. (2008). Model diagnosis for parametric regression in high dimensional spaces. Biometrika. 95. 1-17.

\item Stute, W. and Zhu, L. X. (2002). Model checks for generalized linear models. {\it Scandinavian Journal of Statistics}. {\bf 29}, 535-545.

\item Stute, W. and Zhu, L. X. (2005). Nonparametric checks for single-index models, { \it The Annals of Statistics}, {\bf 33}, 1048-1083.

\item Van Keilegom, I., Gonz\'{a}les-Manteiga, W. and S\'{a}nchez Sellero, C. (2008). Goodness-of-fit tests in parametric regression based on the estimation of the error distribution. {\it TEST}, {\bf 17}, 401-415.

 \item  Wong, H.L., Fang, K.T. and  Zhu,  Lixing (1995). A test for
multivariate normality based on sample entropy and projection
pursuit. {\it  J. of Statistical planning and inference.} 45,
373-385.

\item Xia, Y. C. (2006). Asymptotic distributions for two estimators of the single index model. {\it Econometric Theory}, {\bf 22}, 1112-1137.

\item Xia, Y. C. (2009). Model checking in regression via dimension reduction. {\it Biometrika}, {\bf 96}, 133-148.

\item Xia, Y. C., Tong, H., Li, W. K. and Zhu, L. X. (2002). An adaptive estimation of dimension reduction space (with discussion). {\it Journal of the Royal Statistical Society: Series B}, {\bf 64}, 363-410.

\item Xia, Q., Xu, W. L. and Zhu. L. X. (2014). Consistently determining the number of factors in multivariate volatility modelling. {\it Statistica Sinica}, accepted.

\item Zhang, C. and Dette, H. (2004). A power comparison between nonparametric regression tests. {\it Statistics $\&$ Probability Letters}, {\bf 66}, 289-301.

\item Zheng, J. X. (1996). A consistent test of functional form via nonparametric estimation techniques. {\it Journal of Econometrics}, {\bf 75}, 263-289.

\item Zhu, L. X. (2003). Model checking of dimension-reduction type for regression. {\it Statistica Sinica}, {\bf 13}, 283-296.

\item Zhu, L. X. and An, H. Z. (1992). A nonlinearity test in regression models. {\it Journal of Mathematics}, {\bf 12}, 391-397. (in Chinese)

\item Zhu, L. X. and Fang, K. T. (1996). Asymptotics for the kernel estimates of sliced inverse regression. {\it Annals of Statistics}, {\bf 24}, 1053-1067.
\item Zhu, L.X., and Li R.
(1998). Dimension-reduction type test for linearity of a stochastic model.
{\it Acta Math. Appli. Sinica}, {\bf 14}, 165 - 175.
\item Zhu, L. X. and Ng, K. W. (1995). Asymptotics for sliced inverse regression. {\it Statistica Sinica}, {\bf 5}, 727-736.

\item Zhu, L. P., Zhu, L.X. , Ferr\'{e}, L. and Wang, T. (2010a). Sufficient dimension reduction through discretization-expectation estimation. {\it Biometrika}, {\bf 97}, 295-304.

\item Zhu, L. P., Zhu, L. X. and Feng, Z. H. (2010b). Dimension reduction in regressions through cumulative slicing estimation. {\it Journal of the American Statistical Association} , {\bf 105}, 1455-1466.

\item Zhu, L. X. and Ng, K. W. (1995). Asymptotics for sliced inverse regression. {\it Statistica Sinica}, {\bf 5}, 727-736.

\item Zhu, X. H., Guo, X. and Zhu, L. X. (2014). Model checking for generalized partially linear models: a dimension reduction approach. {\it Working paper.}

\end{description}

\
\newpage

\begin{figure}
  \centering
  \includegraphics[width=16cm,height=12cm]{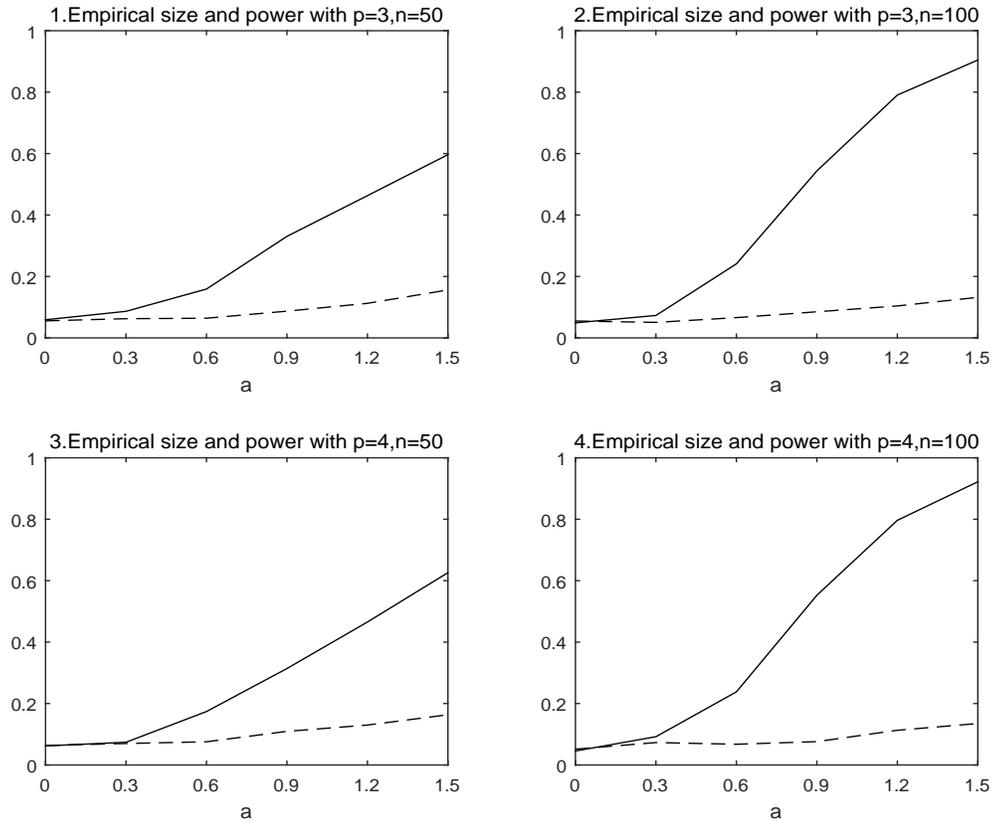}
  \caption{The empirical sizes and powers of $T_n^{SZ}$ and $W_n^2$ in Example 1. The dash and solid line denote the results of $T_n^{SZ}$ and $W_n^2$ respectively.}\label{Figure 1}
\end{figure}

\begin{table}[ht!]\caption{Empirical sizes and powers of $T_n^{GWZ}$, $W_n^2$ and $T_n^{SGP}$ for $H_0$ vs. $H_{11},H_{12}$ and $H_{13}$ in Example 2.}
\centering
{\small\scriptsize\hspace{12.5cm}
\renewcommand{\arraystretch}{1}\tabcolsep 0.5cm
\begin{tabular}{cccccccc}
\hline
&  \multicolumn{1}{c}{a}& \multicolumn{2}{c}{$T_n^{GWZ}$} & \multicolumn{2}{c}{$W_n^2$} &  \multicolumn{2}{c}{$T_n^{SGP}$} \\
&&\multicolumn{1}{c}{n=50}&\multicolumn{1}{c}{n=100}&\multicolumn{1}{c}{n=50}&\multicolumn{1}{c}{n=100}&\multicolumn{1}{c}{n=50}&\multicolumn{1}{c}{n=100} \\
\hline
$H_{11},X\sim N(0,I_p)$  &0.0     &0.0445  &0.0480   &0.0470  &0.0525   &0.0640 &0.0650\\
                     &0.2     &0.0830  &0.1490   &0.0635  &0.0950   &0.1050 &0.1320\\
                     &0.4     &0.1915  &0.4595   &0.1180  &0.2155   &0.1710 &0.3260\\
                     &0.6     &0.4245  &0.8115   &0.2005  &0.4245   &0.2930 &0.5680\\
                     &0.8     &0.6025  &0.9590   &0.3090  &0.6480   &0.4600 &0.8010\\
                     &1.0     &0.7590  &0.9915   &0.4170  &0.8285   &0.5830 &0.9060\\
\hline
$H_{11},X\sim N(0,\Sigma)$ &0.0       &0.0470  &0.0460   &0.0465 &0.0505   &0.0760 &0.0680\\
                       &0.2       &0.0655  &0.1090   &0.0500 &0.0495   &0.0910 &0.0650\\
                       &0.4       &0.1350  &0.3595   &0.0485 &0.0590   &0.0790 &0.0960\\
                       &0.6       &0.2645  &0.6870   &0.0645 &0.0935   &0.0970 &0.1370\\
                       &0.8       &0.4065  &0.8905   &0.0640 &0.1035   &0.1100 &0.1700\\
                       &1.0       &0.5580  &0.9780   &0.0840 &0.1650   &0.1300 &0.2280\\
\hline
$H_{12},X\sim N(0,I_p)$   &0.0          &0.0450  &0.0515   &0.0495 &0.0535   &0.0590 &0.0590\\
                      &0.2          &0.0530  &0.0680   &0.0750 &0.1220   &0.0930 &0.1190\\
                      &0.4          &0.0965  &0.1550   &0.1865 &0.3430   &0.2180 &0.3260\\
                      &0.6          &0.1670  &0.3145   &0.3505 &0.6565   &0.3610 &0.6020\\
                      &0.8          &0.2595  &0.5400   &0.5320 &0.8705   &0.5550 &0.8420\\
                      &1.0          &0.3685  &0.7535   &0.7085 &0.9655   &0.7170 &0.9570\\
\hline
$H_{12},X\sim N(0,\Sigma)$   &0.0        &0.0520  &0.0540   &0.0525  &0.0510   &0.0760 &0.0680\\
                         &0.2        &0.0955  &0.1675   &0.1705  &0.4230   &0.2020 &0.3420\\
                         &0.4        &0.2465  &0.5385   &0.5050  &0.8770   &0.4370 &0.7460\\
                         &0.6        &0.4510  &0.8520   &0.7330  &0.9900   &0.6670 &0.9130\\
                         &0.8        &0.6455  &0.9670   &0.8780  &0.9995   &0.7980 &0.9510\\
                         &1.0        &0.7940  &0.9935   &0.9550  &1.0000   &0.8980 &0.9600\\
\hline
\end{tabular}
}
\end{table}

\newpage
\begin{table}[ht!]\caption{Empirical sizes and powers of $T_n^{GWZ}$, $W_n^2$ and $T_n^{SGP}$ for $H_0$ vs. $H_{11},H_{12}$ and $H_{13}$ in Example 2.}
\centering
{\small\scriptsize\hspace{12.5cm}
\renewcommand{\arraystretch}{1}\tabcolsep 0.5cm
\begin{tabular}{cccccccc}
\hline
&  \multicolumn{1}{c}{a}& \multicolumn{2}{c}{$T_n^{GWZ}$} & \multicolumn{2}{c}{$W_n^2$} &  \multicolumn{2}{c}{$T_n^{SGP}$} \\
&&\multicolumn{1}{c}{n=50}&\multicolumn{1}{c}{n=100}&\multicolumn{1}{c}{n=50}&\multicolumn{1}{c}{n=100}&\multicolumn{1}{c}{n=50}&\multicolumn{1}{c}{n=100} \\
\hline
$H_{13},X\sim N(0,I_p)$      &0.0           &0.0450  &0.0500    &0.0490  &0.0500    &0.0790 &0.0650\\
                         &0.2           &0.0540  &0.0735    &0.1075  &0.1610    &0.1280 &0.1640\\
                         &0.4           &0.0990  &0.2030    &0.2605  &0.5250    &0.2100 &0.3870\\
                         &0.6           &0.1905  &0.4590    &0.4610  &0.8350    &0.3970 &0.6900\\
                         &0.8           &0.3365  &0.7550    &0.6625  &0.9575    &0.5520 &0.8660\\
                         &1.0           &0.4830  &0.9120    &0.7925  &0.9940    &0.7120 &0.9620\\
\hline
$H_{13},X\sim N(0,\Sigma)$  &0.0                  &0.0495 &0.0480   &0.0500   &0.0470   &0.0710 &0.0740\\
                        &0.2                  &0.1385 &0.2640   &0.3565   &0.7110   &0.3050 &0.5380\\
                        &0.4                  &0.4490 &0.8575   &0.7930   &0.9920   &0.6910 &0.9610\\
                        &0.6                  &0.7750 &0.9935   &0.9455   &0.9995   &0.8970 &0.9970\\
                        &0.8                  &0.9005 &0.9995   &0.9790   &1.0000   &0.9700 &1.0000\\
                        &1.0                  &0.9525 &1.0000   &0.9925   &1.0000   &0.9860 &1.0000\\
\hline
\end{tabular}
}
\end{table}

%

\newpage
\begin{table}[ht!]\caption{Empirical sizes and powers of $T_n^{ZH}$ and $W_n^2$ for $H_0$ vs. $H_{31}$ in Example 3.}
\centering
{\small\scriptsize\hspace{12.5cm}
\renewcommand{\arraystretch}{1}\tabcolsep 0.5cm
\begin{tabular}{cccccccc}
\hline
&  \multicolumn{1}{c}{a}& \multicolumn{2}{c}{$T_n^{ZH}$} & \multicolumn{2}{c}{$W_n^2$}  \\
&&\multicolumn{1}{c}{n=50}&\multicolumn{1}{c}{n=100}&\multicolumn{1}{c}{n=50}&\multicolumn{1}{c}{n=100} \\
\hline
$H_{31},X\sim N(0,I_p),p=2$        &0.0          &0.0345  &0.0430  &0.0465 &0.0500\\
                                   &0.2          &0.0820  &0.1505  &0.2095 &0.4375\\
                                   &0.4          &0.3020  &0.6170  &0.6240 &0.9210\\
                                   &0.6          &0.6180  &0.9440  &0.8615 &0.9925\\
                                   &0.8          &0.8410  &0.9930  &0.9445 &1.0000\\
                                   &1.0          &0.9345  &0.9995  &0.9885 &1.0000\\
\hline
$H_{31},X\sim N(0,I_p),p=8$      &0.0                 &0.0265 &0.0295   &0.0500   &0.0450\\
                                 &0.2                 &0.0260 &0.0475   &0.2095   &0.4190\\
                                 &0.4                 &0.0360 &0.0850   &0.5770   &0.9020\\
                                 &0.6                 &0.0765 &0.1640   &0.8100   &0.9935\\
                                 &0.8                 &0.1145 &0.2600   &0.9260   &0.9980\\
                                 &1.0                 &0.1635 &0.3805   &0.9560   &1.0000\\
\hline
$H_{31},X\sim N(0,\Sigma),p=2$   &0.0                  &0.0315  &0.0390   &0.0520  &0.0475\\
                                 &0.2                  &0.0930  &0.1565   &0.2335  &0.4660\\
                                 &0.4                  &0.3250  &0.6530   &0.6275  &0.9305\\
                                 &0.6                  &0.6515  &0.9550   &0.8690  &0.9955\\
                                 &0.8                  &0.8740  &0.9985   &0.9510  &1.0000\\
                                 &1.0                  &0.9550  &1.0000   &0.9775  &1.0000\\

\hline
$H_{31},X\sim N(0,\Sigma),p=8$   &0.0               &0.0185  &0.0350   &0.0465  &0.0530\\
                                 &0.2               &0.0695  &0.1495   &0.5565  &0.9055\\
                                 &0.4               &0.2045  &0.4365   &0.9330  &1.0000\\
                                 &0.6               &0.3370  &0.7320   &0.9835  &1.0000\\
                                 &0.8               &0.4740  &0.8580   &0.9930  &1.0000\\
                                 &1.0               &0.5545  &0.9200   &0.9970  &1.0000\\
\hline
\end{tabular}
}
\end{table}

\newpage
\begin{table}[ht!]\caption{Empirical sizes and powers of $T_n^{ZH}$ and $W_n^2$ for $H_0$ vs. $H_{32}$ in Example 3.}
\centering
{\small\scriptsize\hspace{12.5cm}
\renewcommand{\arraystretch}{1}\tabcolsep 0.5cm
\begin{tabular}{cccccccc}
\hline
&  \multicolumn{1}{c}{a}& \multicolumn{2}{c}{$T_n^{ZH}$} & \multicolumn{2}{c}{$W_n^2$}  \\
&&\multicolumn{1}{c}{n=50}&\multicolumn{1}{c}{n=100}&\multicolumn{1}{c}{n=50}&\multicolumn{1}{c}{n=100} \\
\hline
$H_{32},X\sim N(0,I_p),p=2$        &0.0          &0.034 &0.0455   &0.0530  &0.0500\\
                                   &0.2          &0.083 &0.1155   &0.1215  &0.2050\\
                                   &0.4          &0.250 &0.4730   &0.3430  &0.6100\\
                                   &0.6          &0.524 &0.8480   &0.6200  &0.9195\\
                                   &0.8          &0.782 &0.9785   &0.8540  &0.9920\\
                                   &1.0          &0.935 &0.9985   &0.9575  &0.9995\\
\hline
$H_{32},X\sim N(0,I_p),p=8$      &0.0                     &0.0215 &0.0265    &0.0550  &0.0480\\
                                 &0.2                     &0.0285 &0.0375    &0.1185  &0.1970\\
                                 &0.4                     &0.0475 &0.0760    &0.3215  &0.5830\\
                                 &0.6                     &0.0650 &0.1550    &0.5690  &0.8910\\
                                 &0.8                     &0.1280 &0.2930    &0.7965  &0.9900\\
                                 &1.0                     &0.1765 &0.4210    &0.9230  &1.0000\\

\hline
$H_{32},X\sim N(0,\Sigma),p=2$   &0.0                 &0.0310  &0.0430   &0.0510  &0.0515\\
                                 &0.2                 &0.0745  &0.1410   &0.1205  &0.1880\\
                                 &0.4                 &0.2545  &0.4900   &0.3190  &0.5955\\
                                 &0.6                 &0.5420  &0.8540   &0.6160  &0.9150\\
                                 &0.8                 &0.8105  &0.9835   &0.8400  &0.9880\\
                                 &1.0                 &0.9420  &0.9990   &0.9480  &0.9995\\
\hline
$H_{32},X\sim N(0,\Sigma),p=8$   &0.0                   &0.0270 &0.0295   &0.0520  &0.0470\\
                                 &0.2                   &0.0235 &0.0395   &0.0885  &0.1430\\
                                 &0.4                   &0.0485 &0.0770   &0.1895  &0.3720\\
                                 &0.6                   &0.0725 &0.1480   &0.3750  &0.7000\\
                                 &0.8                   &0.1145 &0.2845   &0.5705  &0.9015\\
                                 &1.0                   &0.1900 &0.4605   &0.7465  &0.9715\\
\hline
\end{tabular}
}
\end{table}

\begin{figure}
  \centering
  \includegraphics[width=16cm,height=16cm]{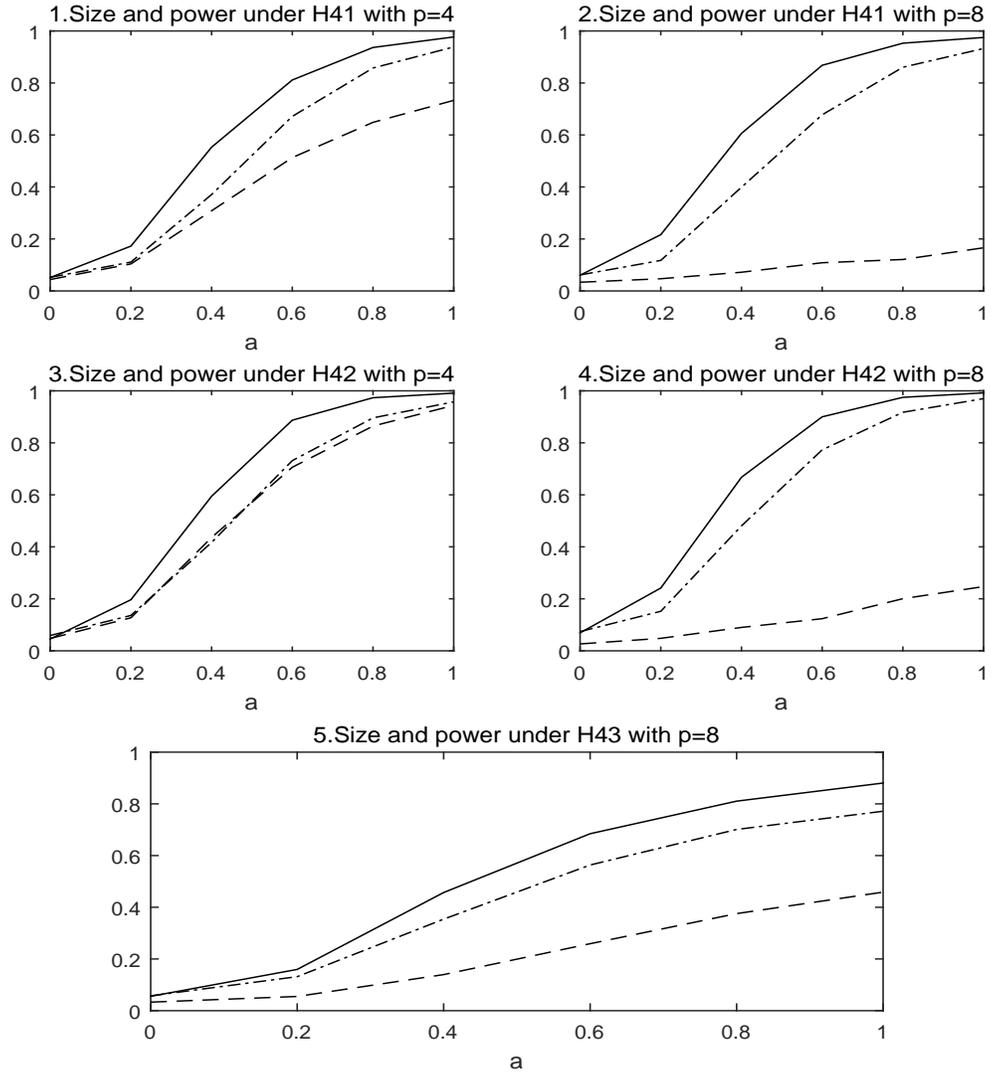}
  \caption{The empirical sizes and powers of $T_n^{ZH}$, $T_n^{GWZ}$and $W_n^2$ in Example~4. The dash, dash-dotted and solid line denote the results of $T_n^{ZH}$, $T_n^{GWZ}$and $W_n^2$ respectively.}\label{Figure 2}
\end{figure}

\begin{figure}
  \centering
  \includegraphics[width=16cm,height=14cm]{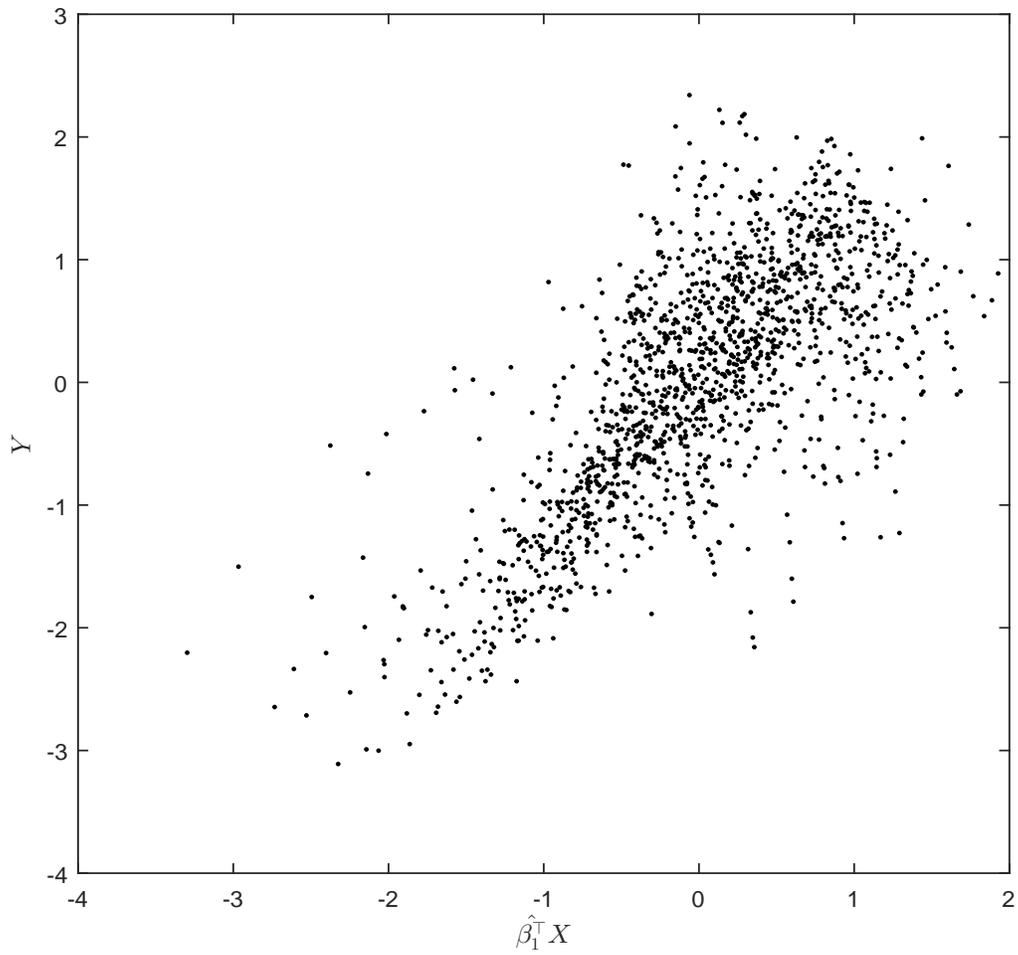}
    \caption{Scatter plot of the response against the $\hat{\beta}_1^{\top}X$ in which the direction $\hat{\beta}_1$ is obtained by DEE.}\label{Figure 3}
\end{figure}

\begin{figure}
  \centering
  \includegraphics[width=16cm,height=14cm]{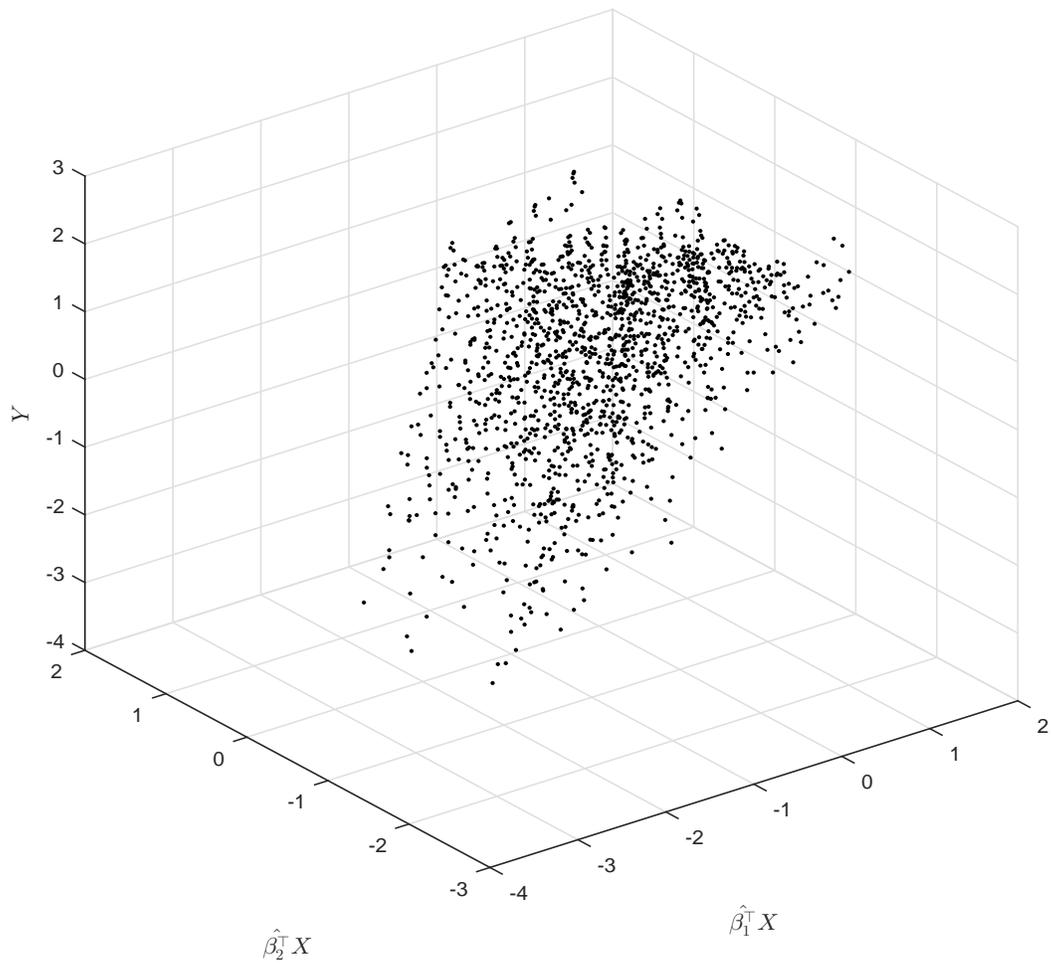}
    \caption{Scatter plot of the response against the $(\hat{\beta}_1^{\top}X, \hat{\beta}_2^{\top}X)$ in which the directions $\hat{\beta}_1$ and $\hat{\beta}_2$ are obtained by DEE.}\label{Figure 4}
\end{figure}


\begin{figure}
  \centering
  \includegraphics[width=16cm,height=14cm]{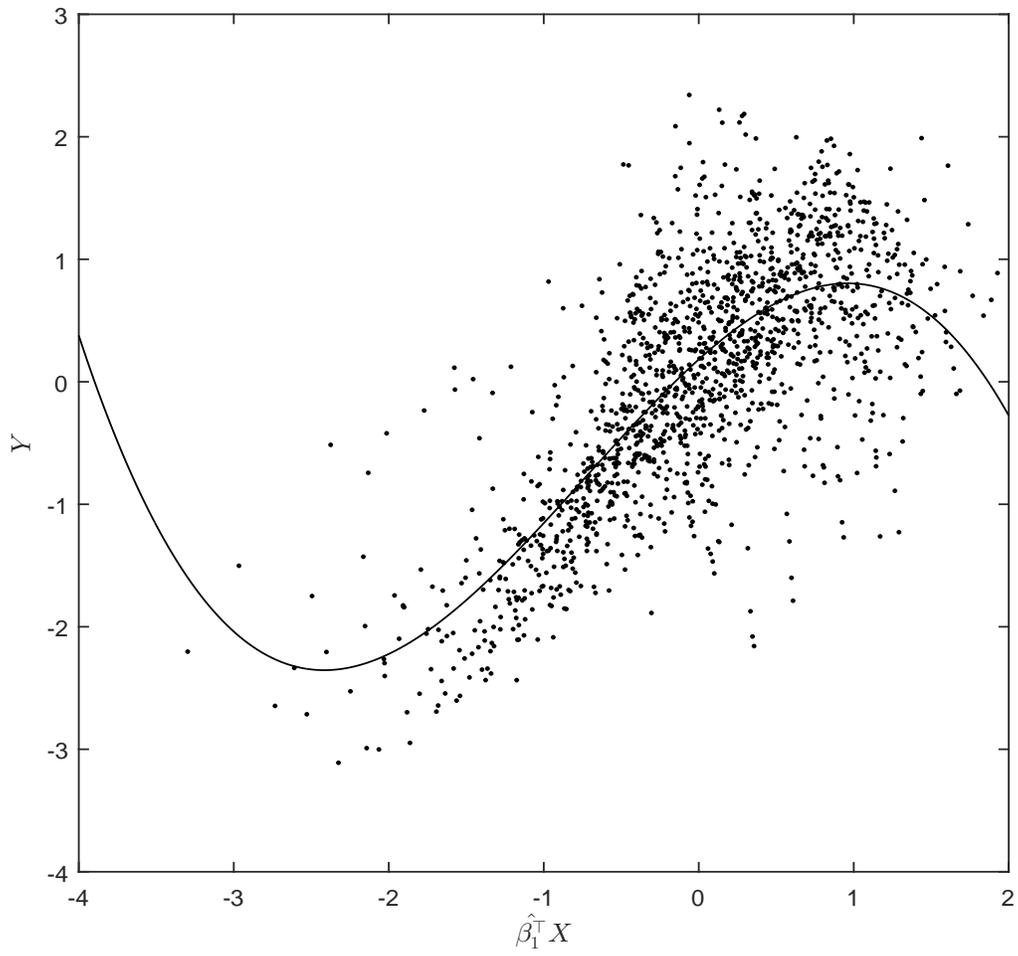}
  \caption{Plot of $Y$ against  $\hat{\beta}_1^{\top}X$ obtained by DEE and the fitted cubic polynomial curve.}\label{Figure 5}
\end{figure}

%

\end{document}